\documentclass[12pt]{article}

\usepackage{amscd}

\usepackage{verbatim}

\usepackage{amssymb}

\usepackage{amsmath}

\newcommand{\be}{\begin{equation}}
\newcommand{\ee}{\end{equation}}
\newcommand{\ben}{\begin{eqnarray}}
\newcommand{\een}{\end{eqnarray}}
\newcommand{\ba}{\begin{align}}
\newcommand{\ea}{\end{align}}
\newcommand{\nn}{\nonumber \\}
\newcommand{\bi}{\begin{itemize}}
\newcommand{\ei}{\end{itemize}}

\newcommand{\lb}{\left (}
\newcommand{\rb}{\right )}
\newcommand{\ltb}{\left [}
\newcommand{\rtb}{\right ]}

\newcommand{\p}{\partial}

\newcommand{\tc}{\tilde{\chi}}
\newcommand{\bc}{\hat{\chi}}

\renewcommand{\theequation}{\thesection.\arabic{equation}}
\begin{document}
\begin{center}

\vspace{24pt} { \large \bf Study of Semiclassical Instability of the Schwarzschild AdS Black Hole in the Large $D$ Limit} \\

\vspace{30pt}

\vspace{30pt}

\vspace{30pt}

{\bf Amruta Sadhu\footnote{sadhuamruta@students.iiserpune.ac.in}}, {\bf Vardarajan
Suneeta\footnote{suneeta@iiserpune.ac.in}}

\vspace{24pt} 
{\em  The Indian Institute of Science Education and Research (IISER),\\
Pune, India - 411008.}

\end{center}
\date{\today}
\bigskip

\begin{center}
{\bf Abstract}
\end{center}

We analyze the semiclassical stability of the Schwarzschild AdS black hole in the Euclidean partition function approach. We perform this computation in the large $D$ limit and focus on scalar perturbations. We obtain the equations for non-spherically symmetric scalar perturbations in a simple form. For a class of perturbations
 stability is demonstrated by the S-deformation method. For some other classes we rule out unstable modes of $\mathcal{O}(D^2)$. We also analyze the spherically symmetric perturbations and demonstrate the appearance of an unstable mode for small black holes in the large $D$ limit. We obtain an expression for the eigenvalue corresponding to the unstable mode to next to leading order in a $1/D$ expansion. This result agrees with a previously obtained numerical bound on this eigenvalue. For cosmological constant zero,  our answer matches a previous result obtained for the corresponding eigenvalue
for the $D$ dimensional Schwarzschild-Tangherlini black hole to next to leading order in a  $1/D$ expansion.

 \setcounter{footnote}{0}
\newpage
\section{Introduction}

In the Euclidean partition function approach, it is well-known from the work of Gross, Perry and Yaffe \cite{gpy} that the
four dimensional Euclidean Schwarzschild instanton has an instability - there exist metric perturbations that
decrease the action. The unstable mode is spherically symmetric, transverse,  traceless and static, and there are no other unstable modes. The analogous
problem for the $D$ dimensional Schwarzschild-Tangherlini black hole was studied in \cite{kol} \footnote{See also \cite{kol1}.} in the limit of large dimension $D$ and an unstable
mode was found. Non-spherically symmetric perturbations of Schwarzschild-Tangherlini black holes in semiclassical gravity were studied in \cite{noneg} and the
black hole was shown to be stable for this class of perturbations in the large $D$ limit. A similar study of semiclassical (in)stability for $D$ dimensional
Schwarzschild Anti-de Sitter (SAdS) black holes has several motivations, in part due to the AdS/CFT correspondence.
Deep connections have also emerged between semiclassical
stability and thermodynamic stability of black holes. In particular, the Hawking-Page phase transition for SAdS black holes prompts the
question of whether a similar transition is  obtained in a study of semiclassical stability of these black holes. Prestidge studied $\ell = 0$ transverse traceless perturbations and showed
numerically that for this class of modes, the large SAdS black hole was semiclassically stable and an unstable mode appeared on decreasing the black hole mass \cite{prestidge}. Prestidge found that the value of the eigenvalue of the unstable mode  was strictly less than $\frac{D^2}{4}$.

In this paper, we study the semiclassical stability of the $D$ dimensional SAdS black holes under a wide class of non-spherically symmetric scalar perturbations.
We simplify the relevant equations using the gauge-invariant formalism of Ishibashi and Kodama. The resultant equations are then studied in the large $D$ limit.
We also analyze the $\ell = 0 $ mode studied by Prestidge and obtain the analogue of the Hawking-Page phase transition \cite{hawkingpage} in semiclassical gravity {\em analytically} in
a large $D$ limit. In particular, we evaluate the eigenvalue of the unstable mode for the small black hole to next to leading order.
The large $D$ limit was used first in \cite{kol}, \cite{kolstring} to obtain the unstable ($\ell=0$) mode  of the Schwarzschild black hole to $\mathcal{O}(1/D)$.
Our result reproduces this answer to this order upon setting the cosmological constant to zero.
The large $D$ limit has been employed as an analytical tool to address problems in perturbation theory of black holes and black branes in \cite{emparan}---\cite{tanabe2} . An effective theory of black holes in a $1/D$ expansion has been proposed and extensively studied in \cite{effectiveemparan}---\cite{min7}. Extension of this work to Gauss Bonnet gravity has been done in \cite{chen}---\cite{bin6}.

The plan of this paper is as follows: In section (\ref{nsp}) we give a summary of the procedure to obtain equations for non-spherically symmetric scalar perturbations for studying semiclassical stability of the SAdS black hole. This involves generalizing previously obtained equations for Schwarzschild black holes to include effects of the non zero cosmological constant. The procedure leads to a set of three equations in terms of variables ($\psi,\phi,\eta)$. The resulting equations for $\psi$ and $\phi$ are coupled and the equation for $\eta$ is decoupled. Appendix (\ref{AppA}) has additional details involved this computation. In section (\ref{etaeqnsec}) we analyze the decoupled equation for $\eta$  using the S-deformation method  to show stability of the SAdS black hole for all $D$. Appendix (\ref{AppB}) contains further details of this analysis.  Section (\ref{largen}) is devoted to a discussion of the large $D$ limit  applied to the SAdS black hole. In section (\ref{sifisec}) we analyze the coupled equations for non-spherically symmetric perturbations $\psi$ and $\phi$ 
. The analysis is carried out using matched asymptotic expansions for various cases in the large $D$ limit where the equations decouple. Appendices (\ref{AppC}) and (\ref{AppD}) contain detailed calculations involving the asymptotic expansions.
 We show that there are no unstable modes for the cases considered. In section (\ref{sp}) we have analysed the spherically symmetric perturbation in the equation using $(1/D)$ expansion. Specifically, we show that there are no unstable modes of $\mathcal{O}(D^2)$. We also perform the analysis for modes of $\mathcal{O}(D)$ and show that there are no unstable modes for large AdS black holes but that the small SAdS black holes have an unstable mode. We calculate the value of this mode to next to leading order. All the results are discussed in detail in section (\ref{results}).

\section{Non-spherically symmetric scalar perturbations}\label{nsp}
The SAdS black hole metric in $D=n+2$ dimensions is given by,
\begin{equation}\label{metric}
g_{\mu\nu}dx^\mu dx^\nu= - f(r) dt^2 + f^{-1} (r)dr^2 +r^2 d\Omega^{2}_{n};
\end{equation}
where $$f (r) = \left(1-\frac{2\Lambda}{n(n+1)} r^2-\frac{b^{n-1}}{r^{n-1}}\right).$$
We can write $\frac{2\Lambda}{n(n+1)}=-\sigma^2$ such that for AdS spacetime with $\Lambda <0$, $\sigma$ is positive. For the case $\Lambda=0$, $\sigma=0$. In this notation,
\begin{equation}
f(r)=\left(1+\sigma^2 r^2-\frac{b^{n-1}}{r^{n-1}}\right)
\end{equation}
The horizon of the SAdS black hole is at $r=r_+$, where $r_+$ is defined by $f(r_+)=0$.

We now proceed with an investigation of semiclassical stability of the SAdS black hole in the large $D$ limit. Following Gross, Perry and Yaffe,\cite{gpy} and Prestidge \cite{prestidge} we study the second variation of the action in the SAdS background. This amounts to studying the eigenvalue problem for the linearized Einstein tensor in the SAdS background. The eigenvalue equation is:
\begin{equation}\label{EEeigen}
2\delta G_{\mu \nu}+2\Lambda h_{\mu\nu}=-\lambda h_{\mu \nu};
\end{equation}
 where the perturbed metric is $\bar{g}_{\mu\nu}=g_{\mu\nu}+h_{\mu\nu}$ and the perturbed Einstein tensor is denoted by $\delta G_{\mu\nu}$. As discussed in \cite{prestidge}, nonconformal modes with $\lambda>0$ decrease the action from the background value and lead to an instability in the Euclidean partition function approach.  Thus, unstable modes correspond to perturbations with $\lambda >0$ that are normalizable at the horizon and infinity.

We want to write the eigenvalue equation (\ref{EEeigen}) in terms of gauge invariant variables proposed by Ishibashi and Kodama (IK), \cite{rev}, \cite{iks},\cite{vac}.
Following IK, we make an ansatz for scalar perturbations of the black hole metric (\ref{metric}) as:
\begin{align}
h_{ab}=f_{ab}S \qquad h_{ai}=rf_aS_i \qquad h_{ij}=2r^2(H_L\gamma_{ij}S+H_TS_{ij}).
\end{align}
We use indices $a,b$ to denote indices from the set $r,t$ and indices $i,j$ denote coordinates on sphere. For static perturbations, the functions $(f_{ab},f_a,H_T,H_L)$ are functions of the radial coordinate $r$.
$S, S_i$ and $S_{ij}$ are scalar harmonics satisfying
\begin{align*}
&(\hat{\Delta}+k^2)S=0 \qquad S_i=-\frac{1}{k}\hat{D}_iS \qquad \hat{D}_iS^i=kS \\
& S_{ij}=\frac{1}{k^2}\hat{D}_i\hat{D}_jS+\frac{1}{n}\gamma_{ij}S \qquad S^i_i=0
\end{align*}
$k^2=\ell(\ell+n-1)$ and $\ell=0,1,2...$.

In \cite{iks} \cite{vac}, the gauge invariant variables are defined for modes with $\ell \geq 2$. The cases $\ell=0,1$ are dealt with separately.  For non-spherically symmetric perturbations, we will only consider $\ell\geq 2$. The modes with $k^2 = n$ (i.e., $\ell=1$) are exceptional modes, in the sense that the construction of gauge-invariant variables is not possible in this case (for details, see \cite{iks}). We will follow a different prescription by Prestidge \cite{prestidge} for the spherically symmetric $\ell=0$ mode. For perturbations with $\ell\geq 2$, we first define
\begin{equation}
X_a=\frac{r}{k}\biggl(f_a+\frac{r}{k}D_aH_T\biggr).
\end{equation}
In terms of $X_a$, the gauge invariant variables are
\begin{align}\label{Fab}
& F_{ab}=f_{ab}+D_aX_b+D_bX_a\\
& F=H_L+\frac{1}{n}H_T+\frac{1}{r}D^aX_a
\end{align}
To simplify the equations further, following IK \cite{vac}, we construct three functions from $F_{ab}, F$;
\begin{align}\label{Sca-WYZ}
 W=r^{n-2}(F^t_t -2F) \qquad Y=r^{n-2}(F^r_r -2F) \qquad Z=r^{n-2}F^r_t.
\end{align}
The goal is to write the equations (\ref{EEeigen}) in terms of these variables. For classical perturbations where $\lambda=0$, the equation $2\delta G_{\mu \nu}+2\Lambda h_{\mu\nu} =0$ has been given in \cite{vac} completely in terms of $W,Y$ and $Z$.  This is done by first writing the equations $2\delta G_{\mu \nu}+2\Lambda h_{\mu\nu}=0$ in terms of $F_{ab},F$ and then writing them in terms of $W,Y$ and $Z$
. We do this process for our eigenvalue equations for semiclassical perturbations with nonzero $\lambda$. We first write the equations in terms of $(F_{ab},F)$. The traceless part of the equation  $2\delta G_{ij}+2\Lambda h_{ij}=-\lambda h_{ij}$ becomes,

\begin{equation}\label{EEeigen-ij}
F_a^a+2(n-2)F=2\lambda \frac{r^2}{k^2}H_T
\end{equation}
The left hand side of the above equation is written completely in terms of the gauge invariant variables but the right hand side has an extra term proportional to $\lambda H_T$.  This term corresponds to the right hand side in $2\delta G_{ij}+2\Lambda h_{ij}=-\lambda h_{ij}$, and prevents the equation from being completely in terms of gauge invariant variables. We see that there will be such terms on the right hand sides of all the eigenvalue equations (eg equation for $\delta G_{rr}$ will have terms proportional to $f_{rr}$ in the right hand side). We now write the equations in terms of $W,Y$ and $Z$. From (\ref{Sca-WYF}) and (\ref{EEeigen-ij}), we get
\begin{equation}\label{Sca-WYF}
W+Y+2nF=2\lambda \frac{r^2}{k^2}H_T.
\end{equation}
We see that the term proportional to $H_T$ cannot be combined with other terms as a gauge invariant variable. All these extra terms in the eigenvalue equations of the form $\lambda h_{ij}$ contain various components of the gauge invariant variables $ W,Y$ and $Z$. In addition to the $-\lambda h_{ij}$ terms, we get extra terms proportional to $H_T$ from the left hand side $2\delta G_{ij}+2\Lambda h_{ij}$.  This is due to the fact that the conversion from the initial gauge invariant variables $F_{ab},F$ to $(W,Y,Z)$ variables is done using the equation (\ref{Sca-WYF}), which has extra factors of $H_T$. To get equations that are completely in terms of $W,Y$ and $Z$, we take appropriate combinations of the eigenvalue equations so that the terms of the form $\lambda h_{ij}$  combine to give a gauge invariant variable. We see that these combinations which make the right hand side gauge invariant also lead to cancellation of all the extra $H_T$ terms (see appendix A for details). For our final equations, we define new variables from $W,Y,Z$ \footnote{In \cite{noneg}, these variables are first defined as $\hat{\psi}(r,t),\hat{\phi}(r,t)$ and $\hat{\eta}(r,t)$ and then by modal decompositions made $\hat{\psi}(r,t)=\psi(r)e^{i\omega t}$ etc. As we are concerned with only static perturbations in this paper, we skip this extra step.}.
\begin{align}\label{Sca-def-sifi}
\psi= \frac{f^{1/2}}{r^{(n-4)/2}} W \qquad \phi = \frac{f^{1/2}}{r^{n/2}} Y \qquad \eta = \frac{1}{r^{(n-2)/2}f^{1/2}}Z ;
\end{align}
The equation for $\phi$ is obtained by combining the eigenvalue equations corresponding to $\delta G_{rr}$, $\delta G_{ii}$ , $\delta G_{ri}$ and $\delta G_{ti}$. Similarly, the equation for $\psi$ is obtained by taking suitable combinations of equations corresponding to $\delta G_{rr}$,  $\delta G_{ii}$ , $\delta G_{ri}$ and $\delta G_{ti}$. The equation for $\eta$ is obtained by combining $\delta G_{rt}$, $\delta G_{ri}$ and $\delta G_{ti}$.
  The coupled equations for static scalar perturbations, in terms of variables $ \psi(r),  \phi(r)$ and $ \eta(r)$, are given below.
 \begin{align}\label{psi}
&-\frac{d^2\psi}{dr^2}+\bigg[\frac{n^3-2n^2+8n-8}{4nr^2}+\frac{f'^2}{4f^2}-\frac{(n^2+2n-4)}{2n}\frac{f'}{fr}-\frac{f''}{2f} \nonumber\\
&+\frac{2(n^2-1)\sigma^2}{n f}-\frac{2(n-1)}{nr^2f}+\frac{k^2}{fr^2}+\frac{\lambda}{f}\bigg]\psi = \nonumber\\
&\left[\frac{2(n-1)}{nf}+\frac{2}{n}-\frac{n+2}{n}\frac{rf'}{f}-\frac{r^2 f''}{f}+\frac{f'^2r^2}{2f^2}+\frac{2(n+1)}{n}\frac{\sigma^2r^2}{f}\right]\phi
\end{align}
\begin{align}
\label{phi}&-\frac{d^2\phi}{dr^2}+\bigg[\frac{n^3-2n^2+8n-8}{4nr^2}+\frac{f'^2}{4f^2}-
\frac{(n^2+2n-4)}{2n}\frac{f'}{fr}-\frac{f''}{2f} \nonumber \\
&+\frac{2(n^2-1)\sigma^2}{n f}-\frac{2(n-1)}{nr^2f}+\frac{k^2}{fr^2}+\frac{\lambda}{f}\bigg]\phi = \nonumber \\
&\left[\frac{2(n-1)}{nr^4f}-\frac{2(n-1)}{nr^4}-\frac{2-n}{nr^3}\frac{f'}{f}-\frac{f''}{r^2f}+\frac{f'^2}{2f^2r^2}+\frac{2(n+1)}{n}\frac{\sigma^2}{r^2f}\right]\psi
\end{align}
\begin{align}
\label{eta}&-\frac{d^2\eta}{dr^2}+\left[\frac{n^2-2n}{4r^2}-\frac{(n+2)f'}{2rf}+\frac{3f'^2}{4f^2}-\frac{3f''}{2f}+\frac{2(n+1)\sigma^2}{f} +\frac{k^2}{fr^2}+\frac{\lambda}{f}\right]\eta =0
\end{align}

We want to find solutions for these equations with $\lambda >0$ that are normalizable at both the horizon and infinity.
The $\eta$ equation decouples for the static perturbations. We shall employ two different strategies for analyzing the equations for $\eta$ and  the coupled equations for $(\phi, \psi)$.

\section{The $\eta$ equation}\label{etaeqnsec}
We shall show the stability of the $\eta$ equation using the S-deformation argument described in \cite{rev},\cite{wald}.
Let us define $\xi=f^{-\frac{1}{2}}\eta$. The equation for $\xi$ can be written in Schrodinger form as
\begin{equation}\label{eta-potentialform}
-\frac{d^2\xi}{dr_*^2}+V\xi=0
\end{equation}
where
\begin{equation}\label{xi-potential}
V=\frac{(n^2-2n)}{4r^2}f^2-\frac{(n+2)}{2}\frac{f'f}{r}+f'^2-2f''f+\frac{k^2}{r^2}f+\lambda f+2(n+1)\sigma^2f.
\end{equation}
The coordinate $r_*$ defined by $dr_*=\frac{dr}{f(r)}$ is the tortoise coordinate. Here $-\infty \leq r_*\leq 0$. Let us assume $\lambda$ to be positive. Let us multiply the equation (\ref{eta-potentialform}) by the complex conjugate of $\xi$ and integrate the equation over the range of $r_*$. The equation (\ref{eta-potentialform}) then becomes
\begin{equation}\label{eta-integratedform}
\left.-\xi^*\frac{d\xi}{dr_*}\right\vert_{-\infty}^0+\int_{-\infty}^0\left(\left\vert\frac{d\xi}{dr_*}\right\vert^2+V(r)|\xi|^2 \right) dr_*=0.
\end{equation}

We want the perturbation $\xi$ which vanishes at both the boundaries. By looking at the equation for $\xi$, we see that one can find solutions which satisfy these boundary conditions (see Appendix B for details).
For normalizable solutions, the boundary term in the equation (\ref{eta-integratedform}) is zero. 
However from the equation (\ref{eta-integratedform}), it is obvious that there are no normalizable solutions if $V(r)>0$. 

Even if $V(r)$ has a small region where it is negative,in some cases, the S-deformation method can be used to show stability. Here we give a brief outline of the method. As shown in \cite{rev}, \cite{wald}  we define a new operator with an arbitrary continuous function $S(r_*)$
\begin{equation}
D=\frac{d}{dr_*}+S(r_*).
\end{equation}

Given these conditions, for normalizable solutions to the equation (\ref{eta-potentialform}), we can write
\begin{equation}\label{xi-Sintegrated}
-\ltb \xi^*\frac{d\xi}{dr_*}+S|\xi|^2\rtb_{-\infty}^0 + \int_{-\infty}^0\left(\left\vert\frac{d\xi}{dr_*}+S \xi\right\vert^2+\lb V(r)+\frac{dS}{dr_*}-S^2\rb |\xi|^2 \right)dr_*=0
\end{equation}

We assume $S$ is such that the boundary terms vanish for the normalizable solutions $\xi$.
To show stability it suffices to show that the deformed potential
\begin{equation}\label{xi-def-pot}
\tilde{V}= V(r)+\frac{dS}{dr_*}-S^2
\end{equation}

is positive for the aforementioned boundary conditions. This can be done by choosing $S$ appropriately. In our case we find
\begin{equation}\label{S-deformation}
S=f'(r)=2\sigma^2 r +(n-1)\frac{b^{n-1}}{r^{n}}.
\end{equation}

The chosen $S$ is such that the all the requirements of this method are obeyed. As $f(r) \geq 0$ and $f'(r) >0$ the deformed potential $\tilde{V}>0$.
\begin{equation}
\tilde{V}=\frac{(n^2-2n)}{4r^2}f^2+\frac{(n-2)}{2}\frac{f'f}{r}+\frac{k^2}{r^2}f+\lambda f
\end{equation}

This argument is valid for all values of $n$ and is independent of the black hole parameters. The argument relies on the assumption $\lambda>0$. This implies there are no normalizable solutions to $\eta$ for both small and large black holes with $\lambda>0$.

This analysis is difficult to extend to the coupled equations for $\phi,\psi$ \footnote{See \cite{kimuracoupled} for extending the S deformation method to coupled equations.}. Here we have to resort to the large $n$ limit in which these equations decouple.
\section{The Large $n$ Limit}\label{largen}

Usage of the large $n$ limit an an analytical tool was first proposed by Kol in \cite{kol} and Emparan et al in \cite{emparan}. In this limit, the function $f(r)=1+\sigma^2r^2-\frac{b^{n-1}}{r^{n-1}}$ is steeply increasing in the region close to the black hole horizon $r_+$. 
 This creates two distinct regions in the black hole spacetime, near the black hole where the $f(r)$ rapidly increases and far from the black hole where $f(r)$ can be approximated to that of AdS. In the large $n$ limit, these two regions share an overlap region. This construction allows us to use techniques of matched asymptotic expansions in our equations.

Following \cite{emparan} \cite{qnmadsemparan}, we define a \textit{near region} where $r-r_+\ll r_+$. Far away from the horizon $f(r)\approx 1+\sigma^2r^2$.
 This allows us to define two regions in the space time as follows.
\begin{align*}
&\text{Near region} \hspace{2cm} r-r_+ \ll r_+\\
&\text{Far region} \hspace{2cm} r-r_+ \gg \frac{r_+}{n}
\end{align*}

with an overlap region  given by $$\frac{r_+}{n} \ll r-r_+\ll r_+$$.

We shall introduce a new coordinate $R=\frac{r^{n-1}}{b^{n-1}}$. In terms of $R$, the horizon is situated at $R_h$, given by
\begin{equation}\label{Rh}
R_h=\left(\frac{r_+}{b}\right)^{n-1}=\frac{1}{1+\sigma^2r_+^2}
\end{equation}

We can write $r$ in the near region $r-r_+\ll r_+$ in terms of $R$ by  Taylor expanding $r=bR^{1/(n-1)}$ around $R_h$, as (\ref{rR1}). We do not take any large $n$ limit to arrive at this expression.
\begin{equation}\label{rR1}
r=r_+\ltb 1+\ln\lb\frac{R}{R_h}\rb \frac{1}{n-1}+\ln\lb\frac{R}{R_h}\rb^2 \frac{1}{2(n-1)^2}+\dots\rtb
\end{equation}

The expression (\ref{rR1}) is not a $1/n$ expansion as $r_+$ solves the equation $r_+=b(1+\sigma^2r_+^2)^{-\frac{1}{n-1}}$ and is $n$-dependent. Following \cite{qnmadsemparan}, \cite{min4}, in this paper we consider $\sigma r_+ \sim \mathcal{O}(1)$. In the large $n$ limit we can write,
\begin{equation}
r_+=b\exp\ltb\frac{-\ln (1+\sigma^2r_+^2)}{n-1}\rtb\sim b \ltb 1-\frac{\ln (1+\sigma^2r_+^2)}{n-1}+\dots\rtb
\end{equation}

For $\sigma r_+ \sim \mathcal{O}(1)$, the difference $r_+-b\sim \mathcal{O}\lb \frac{1}{n}\rb$. Thus in the leading order we can replace $r_+\approx b$. Incorporating the expansion of $r_+$, (\ref{rR1}) becomes,
\begin{equation}\label{rR}
r=b\ltb 1+\frac{\ln R}{n-1}+\frac{(\ln R)^2}{2(n-1)^2}+\dots\rtb
\end{equation}

We shall use this expansion of $r$ in the following sections.

\section{Solving the $\psi$ and $\phi$ equations}\label{sifisec}

We will now solve the equations for $\psi$ and $\phi$ in this limit in both the near and far region approximation. We then extend their solution to the overlap region and using matched asymptotic expansions, compare the two solutions.	

\subsection{Near region solutions}\label{sifinearsec}

In the near region, we write the equations (\ref{psi}) and (\ref{phi}) in terms of $R$. As we are in the region where $r-r_+\ll r_+$, we replace $r$ by $b$ in the leading order.
We will restrict to the perturbations with $k^2,\lambda \sim \mathcal{O}(n^2)$.  We find that the leading order equations decouple for the variables $H$ and $G$ defined as,
\begin{align*}
H=\psi+\phi b^2 \quad G= \psi-\phi b^2
\end{align*}

The leading order equations for $H$ and $G$ are,
\begin{align}\label{H eqn}
&-\lb 1+\sigma^2b^2-\frac{1}{R}\rb^2 \ltb R^2\frac{d^2H}{dR^2}+R\frac{dH}{dR}\rtb+ \nn
&\ltb -\frac{1}{4R^2}+\frac{1}{4}\lb 1+\sigma^2b^2-\frac{1}{R}\rb^2+\lb \frac{k^2+\lambda b^2}{(n-1)^2}-\frac{1}{R}\rb \lb 1+\sigma^2b^2-\frac{1}{R}\rb\rtb H=0\\
\label{G eqn}&-\lb 1+\sigma^2b^2-\frac{1}{R}\rb^2 \ltb R^2\frac{d^2G}{dR^2}+R\frac{dG}{dR}\rtb+ \nn
&\ltb \frac{3}{4R^2}+\frac{1}{4}\lb 1+\sigma^2b^2-\frac{1}{R}\rb^2+\lb \frac{k^2+\lambda b^2}{(n-1)^2}+\frac{1}{R}\rb \lb 1+\sigma^2b^2-\frac{1}{R}\rb\rtb G=0
\end{align}

Redefining $(1+\sigma^2b^2)R=\tilde{R}$, we see that these equations can be written as hypergeometric equations with singularities  at $\tilde{R}=0,1$ and $\infty$. The horizon is at $\tilde{R}=1$. The solution to $H$ in terms of hypergeometric functions is
\begin{equation}\label{H soln}
H=\tilde{R}(\tilde{R}-1)^{\frac{1}{2}}\ltb C_1 F(p,q,1,1-\tilde{R})+C_2F(p,q,1,1-\tilde{R})\ln(1-\tilde{R})\rtb
\end{equation}

$C_1$ and $C_2$ are arbitrary constants. Let us denote $\frac{k}{(n-1)}= \hat{k}$ and $\frac{\lambda}{(n-1)^2}= \hat{\lambda}$ The constants $p,q$ are
\begin{align*}
p=\frac{3}{2}+\frac{1}{2}\sqrt{1+\frac{4(\hat{k}^2+\hat{\lambda}b^2)}{(1+\sigma^2b^2)}} \quad q=\frac{3}{2}-\frac{1}{2}\sqrt{1+\frac{4(\hat{k}^2+\hat{\lambda}b^2)}{(1+\sigma^2b^2)}}
\end{align*}

We want perturbations to be normalizable at the horizon, $\tilde{R}=1$. Hence we set $C_2=0$. The normalizable solution to (\ref{H eqn}) is,
\begin{equation}
H=C_1\tilde{R}(\tilde{R}-1)^{\frac{1}{2}} F(p,q,1,1-\tilde{R})
\end{equation}

We now want to extend this solution to the overlap region. Using standard formulae for hypergeometric functions, we can write the solution near $\tilde{R}=1$  as a linear combination of hypergeometric functions of the form $F(\alpha,\beta,\gamma, 1/\tilde{R})$.
\begin{align}
H=C_1\tilde{R}(\tilde{R}-1)^{\frac{1}{2}}&\ltb\frac{\Gamma(q-p)}{\Gamma(q)\Gamma(q-2)}\tilde{R}^{-p}F\lb p,p-2,p-q+1,\frac{1}{\tilde{R}}\rb+\right.\nn &\left.\frac{\Gamma(p-q)}{\Gamma(p)\Gamma(p-2)}\tilde{R}^{-q}F\lb q,q-2,q-p+1,\frac{1}{\tilde{R}}\rb\rtb.
\end{align}

Taking $\tilde{R}$ large, the solution for $H$ in the overlap region is,
\begin{equation}\label{H-ovlp}
H=\frac{C_1\Gamma(q-p)}{\Gamma(q)\Gamma(q-2)}\tilde{R}^{-\frac{1}{2}\sqrt{1+\frac{4(\hat{k}^2+\hat{\lambda}b^2)}{(1+\sigma^2b^2)}}} +\frac{C_1\Gamma(p-q)}{\Gamma(p)\Gamma(p-2)}\tilde{R}^{\frac{1}{2}\sqrt{1+\frac{4(\hat{k}^2+\hat{\lambda}b^2)}{(1+\sigma^2b^2)}}}
\end{equation}

Similarly we can solve the $G$ equation (\ref{G eqn}). The normalizable solution at the horizon is,
\begin{equation}\label{G-soln}
G=D_1(\tilde{R}-1)^{\frac{3}{2}}F(p,q,3,1-\tilde{R})
\end{equation}
which extended to the overlap region becomes
\begin{equation}\label{G-ovlp}
G=\frac{D_12\Gamma(q-p)}{\Gamma(q)^2}\tilde{R}^{-\frac{1}{2}\sqrt{1+\frac{4(\hat{k}^2+\hat{\lambda}b^2)}{(1+\sigma^2b^2)}}} +\frac{D_12\Gamma(p-q)}{\Gamma(p)^2}\tilde{R}^{\frac{1}{2}\sqrt{1+\frac{4(\hat{k}^2+\hat{\lambda}b^2)}{(1+\sigma^2b^2)}}}
\end{equation}

For matching with the far region, we will later find expressions for  $\psi$ and $\phi$ by adding and subtracting (\ref{H-ovlp}) and (\ref{G-ovlp}). 

\subsection{Far region solution}\label{sififarsec}

The far region is defined by $r-r_+ \gg \frac{r_+}{n-1}$. In this region $b^{n-1}/r^{n-1}\sim e^{-n\ln r}$ is a small quantity for both $r$ and $n$ large. For asymptotically AdS spacetimes, we can then approximate the function $f(r) \approx 1+\sigma^2r^2$. In the large $n$ limit, the equations (\ref{psi}) and (\ref{phi}) become

\begin{align}\label{psi-far-hyp-eqn}
&-\frac{d^2\psi}{dr^2}+\lb \frac{n^2+4k^2}{4r^2}+\frac{\lambda-k^2\sigma^2}{1+\sigma^2r^2}-\frac{\sigma^2}{(1+\sigma^2r^2)^2}\rb \psi=\lb\frac{2}{n}+\frac{2}{(1+\sigma^2r^2)^2}\rb \phi\\
\label{phi-far-hyp-eqn}
&-\frac{d^2\phi}{dr^2}+\lb \frac{n^2+4k^2}{4r^2}+\frac{\lambda-k^2\sigma^2}{1+\sigma^2r^2}-\frac{\sigma^2}{(1+\sigma^2r^2)^2}\rb \phi=\frac{2\sigma^4}{(1+\sigma^2r^2)^2} \psi
\end{align}

As in the near region, we assume $k^2,\lambda \sim \mathcal{O}(n^2)$.    Looking at the near region solution $H=\psi+b^2\phi$ in (\ref{H-ovlp}) and $G=\psi-b^2\phi$ in (\ref{G-ovlp}), we know that the two solutions have the same leading order dependence on $R\sim r^n$. But we also notice that there is an extra factor of $b^2$ multiplying the solution $\phi$ in the near horizon region where $r\sim b$. In view of this observation, in the far region, we consider the special case when $\psi\sim r^\gamma\phi$. But from the near region solutions to $\phi$ and $\psi$, we know that $\gamma\ll n$. It is easy to see that for $\gamma>2$, the right hand side of (\ref{psi-far-hyp-eqn}) can be neglected for both large $r$ and large $n$. Similarly the coupling terms in (\ref{phi-far-hyp-eqn}) can be neglected for $\gamma\leq 2$ in the large $n$ approximation. First consider the latter case. The normalizable solution of the decoupled $\phi$ equation is given in terms of a hypergeometric function. For large $n$,
\begin{align}\label{phi-far-hyp-soln}
&\phi=d_1 r^{-\frac{n}{2}\sqrt{1+\frac{4\hat{\lambda}}{\sigma^2}}} ~\times \nn &F\lb \frac{1}{4}\lb\sqrt{n^2+4k^2}+\sqrt{n^2+\frac{4\lambda}{\sigma^2}}\rb,\frac{1}{4}\lb\sqrt{n^2+\frac{4\lambda}{\sigma^2}}-\sqrt{n^2+4k^2}\rb,\frac{1}{2}\sqrt{n^2+\frac{4\lambda}{\sigma^2}},-\frac{1}{\sigma^2r^2}\rb
\end{align}

We would like to extend this solution to the overlap region by writing $r$ in terms of $R$ using (\ref{rR}). For this purpose,  we would also have to use the transformation formulae for hypergeometric functions. But the transformation of the hypergeometric function becomes unwieldy as all the parameters of the function become large. We shall hence approximate $f(r)\approx \sigma^2r^2$ for $r\to \infty$ in (\ref{psi}) and (\ref{phi}).  Using this approximation, we get the solution for $\psi,\phi$ as Bessel functions, which can be extended to the overlap region. We note that this approximation is not valid for small black holes ($\sigma r_+<<1$).\footnote{Large black holes are those with $\sigma r_+>1$. We follow the terminology of Hawking and Page here \cite{hawkingpage}.} This is because in the overlap region we cannot approximate $1+\sigma^2r^2 \sim \sigma^2r^2$. Substituting $f(r)$ in (\ref{psi}) and (\ref{phi}).

\begin{align}\label{psi-far-bes-eqn}
&-\frac{d^2\psi}{dr^2}+\ltb \lb \frac{\lambda}{\sigma^2}+\frac{n^2}{4}\rb\frac{1}{r^2}+\frac{k^2}{\sigma^2r^4}\rtb\psi=\lb 2-\frac{2}{n}\rb \frac{1}{\sigma^2r^2}\phi\\
\label{phi-far-bes-eqn}&-\frac{d^2\phi}{dr^2}+\ltb \lb \frac{\lambda}{\sigma^2}+\frac{n^2}{4}\rb\frac{1}{r^2}+\frac{k^2}{\sigma^2r^4}\rtb\phi=\ltb \frac{2}{r^4}+\lb 2-\frac{2}{n}\rb\frac{1}{\sigma^2r^6}\rtb\psi
\end{align}

Let us again consider $\psi\sim r^\gamma\phi$. For $\gamma>2$, we can neglect the coupling terms in the $\psi$ equation (\ref{psi-far-bes-eqn}). For $\gamma=2$ or $\gamma<1$ we can ignore the coupling terms in the $\phi$ equation (\ref{phi-far-bes-eqn}) in the large $n$ limit. The case $\gamma=1$ can be decoupled only if we keep the leading terms proportional to $r^{-2}$ in the right hand sides of both the equations.  We will examine the case $\gamma=2$ or $\gamma<1$ here. As the right hand sides of both the equations have the same form, leading order solutions for the case $\gamma>2$ can be obtained by replacing $\phi$ by $\psi$ in the following calculation. The leading order equation for $\phi$ is,
\begin{equation}
-\frac{d^2\phi}{dr^2}+\ltb\lb \frac{\lambda}{\sigma^2}+\frac{n^2}{4}\rb\frac{1}{r^2}+\frac{k^2}{\sigma^2r^4}\rtb\phi=0
\end{equation}

The solution to this equation is
\begin{equation}
\phi= \sqrt{r}\ltb c_1 I_\nu\lb \frac{k}{\sigma r}\rb+c_2 K_\nu\lb \frac{k}{\sigma r}\rb \rtb
\end{equation}
where $\nu=\sqrt{\frac{n^2+1}{4}+\frac{\lambda}{\sigma^2}}$.
As $r\to \infty$, the two solutions behave as $I_\nu(\frac{1}{r})\sim r^{-\nu}$ and $K_\nu(\frac{1}{r})\sim r^{\nu}$. As we want the solution to be normalizable at infinity, we choose $c_2=0$. The final solution is
\begin{equation}\label{far-phi-sol}
\phi=c_1\sqrt{r}I_\nu\lb \frac{k}{\sigma r}\rb.
\end{equation}
We now want to extend this solution to the overlap region.
As we are dealing with $k^2,\lambda \sim n^2$, both the argument and order of the modified Bessel function $I_\nu\lb \frac{k}{\sigma r}\rb$ are large in the overlap region. The behaviour of this function is hence given by an uniform asymptotic expansion for modified Bessel functions. Let us define $z$ such that the argument of the Bessel function $\frac{k}{r\sigma}=\nu z$. In terms of $z$, $\phi$ is written as,

\begin{equation}\label{phi-in-z}
\phi=\frac{c_1I_\nu(\nu z)}{\sqrt{z}}=\frac{1}{\sqrt{z}}\frac{e^{\nu\eta}}{(1+z^2)^{1/4}\sqrt{2\pi\nu}}\left[1+\sum_{m=1}^{\infty}\frac{U_m(\tilde{t})}{\nu^m}\right]
\end{equation}

where
\begin{eqnarray}
\eta &=& \sqrt{1+z^2}+\ln\left[\frac{z}{1+\sqrt{1+z^2}}\right]; \nonumber \\
\tilde t &=& \frac{1}{\sqrt{1+z^2}}.
\end{eqnarray}

and $U_m(\tilde t)$ are polynomials in $\tilde t$.
This solution can be extended to the overlap region of the
near and far regions
to get the leading order solution for large $n$. To find the solution in the overlap region,
we write $r$ in terms of $R$ by using (\ref{rR}) which is valid in the
entire near region, and therefore, in particular, in the overlap region. Details of this calculation are given in Appendix C.
The leading order solution in terms of $R$ is,
\begin{equation}\label{phi-ovlp-far}
\phi=c_1 R^{-\frac{1}{2}\sqrt{1+\frac{4(\hat{k}^2+\hat{\lambda}b^2)}{\sigma^2b^2}}}
\end{equation}

We can similarly solve for $\psi$. The leading order solution for $\psi$ in the overlap region is again,

\begin{equation}\label{psi-ovlp-far}
\psi=d_1 R^{-\frac{1}{2}\sqrt{1+\frac{4(\hat{k}^2+\hat{\lambda}b^2)}{\sigma^2b^2}}}
\end{equation}

Let us consider the cases where $\lambda,k^2$ are $\mathcal{O}(n)$ or less. As the coupling terms in the equations (\ref{psi-far-bes-eqn}) and (\ref{phi-far-bes-eqn}) are $\mathcal{O}(1)$, we can decouple the equations using the same logic as the $\lambda,k^2 \sim \mathcal{O}(n^2)$ case. The normalizable solutions for $\phi$ and $\psi$ are $\sim I_\nu \lb \frac{k}{\sigma}r\rb$. The order of the modified Bessel function $I_\nu \lb \frac{k}{\sigma}r\rb$ is $\nu \sim n$ and the argument is $k\sim \sqrt{n}$. As the order and argument are large but of different orders, we cannot use the uniform asymptotic expansions to extend the solutions to the overlap region. For this reason we cannot comment on cases where $\lambda, k^2 \sim \mathcal{O}(n)$ or less.

\subsection{Matching the solutions}\label{sifimatchsec}

From the solutions (\ref{H-ovlp}) and (\ref{G-ovlp}), we can write $\phi$ and $\psi$ in the overlap region as,
\begin{align}\label{phi-ovlp-near}
&\phi= e_1 \tilde{R}^{-\frac{1}{2}\sqrt{1+\frac{4(\hat{k}^2+\hat{\lambda}b^2)}{(1+\sigma^2b^2)}}}+ e_2\tilde{R}^{\frac{1}{2}\sqrt{1+\frac{4(\hat{k}^2+\hat{\lambda}b^2)}{(1+\sigma^2b^2)}}}\\
\label{psi-ovlp-near}&\psi= e_3 \tilde{R}^{-\frac{1}{2}\sqrt{1+\frac{4(\hat{k}^2+\hat{\lambda}b^2)}{(1+\sigma^2b^2)}}}+ e_4\tilde{R}^{\frac{1}{2}\sqrt{1+\frac{4(\hat{k}^2+\hat{\lambda}b^2)}{(1+\sigma^2b^2)}}}
\end{align}

Where
\begin{align}\label{e_is}
e_1=\frac{C_1\Gamma(q-p)}{2b^2\Gamma(q)\Gamma(q-2)}-\frac{D_12\Gamma(q-p)}{2b^2\Gamma(q)^2}\quad e_2=\frac{C_1\Gamma(p-q)}{2b^2\Gamma(p)\Gamma(p-2)}-\frac{D_12\Gamma(p-q)}{2b^2\Gamma(p)^2}\\
e_3=\frac{C_1\Gamma(q-p)}{2\Gamma(q)\Gamma(q-2)}+\frac{D_12\Gamma(q-p)}{2\Gamma(q)^2}\quad e_4=\frac{C_1\Gamma(p-q)}{2\Gamma(p)\Gamma(p-2)}+\frac{D_12\Gamma(p-q)}{2\Gamma(p)^2}
\end{align}

We now want to match these solutions to the solutions obtained from the far region (\ref{phi-ovlp-far}) and (\ref{psi-ovlp-far}). As expected, we see that the exponents in the two solutions become the same only for $\sigma b\gg1$, i.e. for large black holes. Solutions coming from the near region have both a growing and a decaying piece, whereas the solution from far is only decaying. For the matching of solutions, we need the coefficient of the growing piece from the near region to vanish both for $\psi$ and $\phi$ simultaneously. This translates to making both $e_2,e_4=0$. From (\ref{e_is}) we see that by choosing $C_1$ and $D_1$ appropriately, we may only be able to match  either $\psi$ or $\phi$ solution to the far solution, but not both. For both the constants $e_2$ and $e_4$ to vanish simultaneously, we will need the gamma function $\Gamma(p)$ to have a pole, i.e. $p$ to be a nonpositive integer. For the parameter range that we are interested in, namely $k>0, \lambda>0$ and $k^2,\lambda\sim \mathcal{O}(n^2)$, this possibility is ruled out. Hence there are no normalizable modes for SAdS black holes for $\lambda\sim \mathcal{O}(n^2)$.

For other parameter ranges, we cannot apply this analysis as the equations remain coupled in the far region.

\section{Spherically symmetric ($\ell=0$) perturbation}\label{sp}
The equations used in the previous section use IK variables which are only valid for perturbations with $\ell \geq 2$. Although IK provide a prescription to extend these variables to the $\ell=0$  perturbations via gauge fixing, the equations obtained in terms of these variables are unwieldy. A much simpler equation for these perturbations has been obtained in \cite{prestidge}. This equation has been analysed numerically for various $n$ values for SAdS black holes in \cite{prestidge} and analytically for large $n$ for the $\Lambda=0$ black hole in \cite{kol}.  As shown in \cite{gpy} \cite{prestidge}, only the terms operating on the transverse traceless modes can have negative eigenvalues. Following \cite{prestidge} we consider transverse and traceless perturbations $h_{\mu\nu}$ of the form
\begin{equation}
h_\mu^\nu=diag\lb \zeta(r),\chi(r),H_L(r),\dots, H_L(r)\rb
\end{equation}
where $(H_L (r),\dots H_L(r))$ are $n$ terms. For the transverse traceless mode
\begin{equation}\label{traceless}
H_L=-\frac{\zeta+\chi}{n}.
\end{equation}
Subsequently, using $\nabla^\mu h_{\mu\nu}=0$, the relation between $\zeta$ and $\chi$ can be written as
\begin{equation}\label{transverse}
\zeta=\frac{2rf}{rf'-2f}\chi'(r)+\frac{rf'+2(n+1)f}{rf'-2f}\chi(r).
\end{equation}
With this ansatz the relevant perturbation equation reduces to a linear second order ordinary differential equation for $\chi$ \cite{prestidge}.
\begin{align}\label{chi eqn}
-f\chi''+&\frac{2r^2(ff''-f'^2)-rnff'+2(n+2)f^2}{r(rf'-2f)}\chi'\nn&+\frac{r^2f'f''+r[2(n+1)ff''-(n+4)f'^2]+4ff'}{r(rf'-2f)}\chi=-\lambda\chi
\end{align}
As before, we want solutions to this equation that are normalizable at both the horizon and infinity for $\lambda >0$. The singular points of this equation are $0,r_+,\infty$  and $r_s$ where $r_s$ is the solution to $(rf'-2f)=0$. Hence, along with the regularity at both the boundaries, it is desirable for the perturbation to be well-behaved at $r_s$. This implies,  from the equation (\ref{transverse}), at $r_s$
\begin{equation}\label{bdy cond at rs}
\frac{\chi'(r_s)}{\chi(r_s)}=-\frac{rf'+2(n+1)f}{2rf}\bigg|_{r_s}=-\frac{n+2}{r_s}.
\end{equation}

We will later see that both the linearly independent solutions to (\ref{chi eqn}) around the point $r_s$ satisfy this equation, so this is \emph{not} in fact an extra condition on the solution $\chi$.

Even though simpler than the equations for $\ell \geq 2$ case, the equation (\ref{chi eqn}) cannot be solved for the entire range of $r$ analytically. We use the large $n$ limit as an analytical approximation tool to study (in)stability.

\subsection{Far region solution}\label{spfarsec}
We first solve equation (\ref{chi eqn}) in the far region. We proceed by substituting for $f(r)$ in (\ref{chi eqn}) and neglecting the terms of order $r^{-n}$ in the far region. Note that this is just the far region approximation, we have not assumed a large $n$ limit to obtain the following equation:
\begin{equation}\label{chi eqn far}
(1+\sigma^2r^2)\chi''+\frac{(n+2)+(n+6)\sigma^2r^2}{r}\chi'+[(2n+6)\sigma^2-\lambda]\chi=0.
\end{equation}
The solutions of this equation  are,
\begin{align}
\chi=&D_1(\sigma^2r^2)^{-p}F\lb p,p-s+1,p-q+1,-\frac{1}{\sigma^2r^2}\rb+\nn &D_2(\sigma^2r^2)^{-q}F\lb q,q-s+1,q-p+1,-\frac{1}{\sigma^2r^2}\rb
\end{align}
Here $s=\frac{n+3}{2}$ and
\begin{align}
p=\frac{(n+5)}{4}+\frac{1}{2}\sqrt{\frac{(n+1)^2}{4}+\frac{\lambda}{\sigma^2}};\quad q=\frac{(n+5)}{4}-\frac{1}{2}\sqrt{\frac{(n+1)^2}{4}+\frac{\lambda}{\sigma^2}}.
\end{align}
At $r\to \infty$, the hypergeometric function can be can be truncated to its leading term.

\begin{equation}
\chi \approx D_1 r^{-\frac{n}{2}-\frac{n}{2}\sqrt{1+4\frac{\hat{\lambda}}{\sigma^2}}}+D_2 r^{-\frac{n}{2}+\frac{n}{2}\sqrt{1+4\frac{\hat{\lambda}}{\sigma^2}}}
\end{equation}

Here $\hat{\lambda}=\frac{\lambda}{(n-1)^2}$. We want to choose the solution that is normalizable at infinity.
For $\hat{\lambda} >0$, this condition dictates $D_2=0$. In terms of $R$, the leading order solution in the far region for the case $\lambda \sim \mathcal{O}(n^2)$ is
\begin{equation}\label{far soln nd}
\chi\approx D_1 R^{-\frac{1}{2}-\frac{1}{2}\sqrt{1+4\frac{\hat{\lambda}}{\sigma^2}}}
\end{equation}
For the cases $\lambda$ of order $\mathcal{O}(n)$ or lower, the far region solution at the leading order in $n$ is
\begin{equation}\label{far soln dec}
\chi \approx D_1\frac{1}{R}.
\end{equation}

\subsection{Near region solution}\label{spnearsec}
We now turn our attention to the near region.
 As before, to get the equation in the near region, we write  (\ref{chi eqn}) in terms of the $R$ coordinate.
\begin{align}\label{chi eqn near}
\frac{d^2\chi}{dR^2}+&\ltb \frac{2n}{(n-1)R}-\frac{4}{2R-(n+1)}+\frac{2((n-1)(1+\sigma^2r^2)+2\sigma^2r^2)}{(n-1)(R(1+\sigma^2r^2)-1)}\rtb \frac{d\chi}{dR}\nn &\hspace{1.3in}+\frac{2(AB R-C)}{R(2R-(n+1))(R(1+\sigma^2r^2)-1)}\chi=0
\end{align}
Where
\begin{align}
&C=(n-1)(1+\sigma^2r^2)-\frac{\lambda r^2}{2(n-1)}+3+6\sigma^2r^2+\frac{12\sigma^2r^2}{(n-1)}+\frac{8\sigma^2r^2-\lambda r^2}{(n-1)^2}\\
&AB= \frac{-\lambda r^2+(8+2(n-1))\sigma^2r^2}{(n-1)^2}
\end{align}

Even for the leading value of $r \approx b$ in the near horizon region, equation (\ref{chi eqn near}) is a Heun equation with singularities at $0,\frac{1}{1+\sigma^2b^2}, \frac{n+1}{2}$ and $\infty$. Solutions of these type of equations can generically be written as a power series around each singular point. There are no connection formulae between the solutions at two singular points and thus not much can be inferred from them. We notice that in the equation (\ref{chi eqn near}), different terms assume importance   very near the horizon at $\frac{1}{1+\sigma^2b^2}$ where $R\sim \mathcal{O}(1)$ and near the singular point $R=\frac{n+1}{2}$ corresponding to $r_s$ where $R\sim \mathcal{O}(n)$. We will hence proceed by solving the equation in the two regimes, $R \sim \mathcal{O}(1)$ and $R \sim \mathcal{O}(n)$. In both the regimes, we shall employ a $(1/n)$ expansion for $\chi$ and $\lambda$. As both the regimes are still in the original near region $r-r_+\ll r_+$, the full near region solution must be obtained by matching the two solutions. To this end, we propose an overlap region between the two singular points such that $1\ll R\ll n$. We then match the solutions in this new overlap region. We also expect the whole near region solution to match with the far region solution for $R \to \infty$. This two step matching procedure gives us the value of $\lambda$. This matching scheme has been used by Emparan, Suzuki, Tanabe in \cite{est} to obtain decoupled quasinormal modes for Schwarzschild-Tangherlini black holes.

Henceforth we will denote
\begin{equation}
n-1=m.
\end{equation}

The perturbation variable $\chi$ and eigenvalue $\lambda$ are expanded as,
\begin{align}
\chi= \sum_{i=0}^{\infty}\frac{\chi_i}{m^i}; \qquad \lambda= \sum_{i=0}^{\infty}\frac{\lambda_i}{m^i}
\end{align}

Let us first consider the case $\lambda \sim \mathcal{O}(n^2)$.

\subsubsection{The case $\lambda \sim \mathcal{O}(n^2)$}
Very near the horizon, $R \sim \mathcal{O}(1)$. We rewrite the equation (\ref{chi eqn near}) in terms of the coordinate $x=1-(1+\sigma^2b^2)R$ and denote $\chi$ as $\tc$ in this region. In terms of $x$, the horizon lies at $x=0$. In the leading order,
the equation becomes
\begin{equation}\label{chi near hor}
x(1-x)\frac{d^2\tc_0}{dx^2}+(2-4x)\frac{d\tc_0}{dx}-\lb2-\frac{\hat{\lambda} b^2}{(1+\sigma^2b^2)}\rb \tc_0=0
\end{equation}

Here we have written $\hat{\lambda}=\frac{\lambda}{(n-1)^2}$ such that $\hat{\lambda}\sim \mathcal{O}(1)$.  Let us denote $$\delta=\frac{\hat{\lambda} b^2}{(1+\sigma^2b^2)}.$$ Solutions of this equation are
\begin{align}
\tc_0=&C_1F\lb\frac{3}{2}+\frac{\sqrt{1+4\delta}}{2},\frac{3}{2}-\frac{\sqrt{1+4\delta}}{2},2, x\rb+\nn
&C_2\lb F\lb\frac{3}{2}+\frac{\sqrt{1+4\delta}}{2},\frac{3}{2}-\frac{\sqrt{1+4\delta}}{2},2, x\rb\ln x+\frac{a_1}{x}+\sum_{i=0}^{k}c_kx^k\rb;
\end{align}

where $a_1$ and $c_k$ are constants.\footnote{The values of these constants can be found in chapter 15 of Handbook of Mathematical Functions by M Abrahmowitz and I Stegun.} We choose $C_2=0$ as we want $\chi_0$ to be normalizable at $x=0$. In order to compare with the solution at $R\approx m/2$, let us extend this solution to the overlap region within the near region  $1\ll R\ll n$ by taking $x$ large. The far limit of the near horizon solution is
\begin{align}
\tc_0=&C_1\frac{\Gamma(-\sqrt{1+4\delta})x^{-\frac{3}{2}-\frac{\sqrt{1+4\delta}}{2}}}{\Gamma\lb \frac{3}{2}-\frac{\sqrt{1+4\delta}}{2}\rb\Gamma\lb \frac{1}{2}-\frac{\sqrt{1+4\delta}}{2}\rb}F\lb  \frac{3}{2}+\frac{\sqrt{1+4\delta}}{2}, \frac{1}{2}+\frac{\sqrt{1+4\delta}}{2}, 1+\sqrt{1+4\delta},\frac{1}{x}\rb+\nn
&C_1\frac{\Gamma(\sqrt{1+4\delta})x^{-\frac{3}{2}+\frac{\sqrt{1+4\delta}}{2}}}{\Gamma\lb \frac{3}{2}+\frac{\sqrt{1+4\delta}}{2}\rb\Gamma\lb \frac{1}{2}+\frac{\sqrt{1+4\delta}}{2}\rb}F\lb  \frac{3}{2}-\frac{\sqrt{1+4\delta}}{2}, \frac{1}{2}-\frac{\sqrt{1+4\delta}}{2}, 1-\sqrt{1+4\delta},\frac{1}{x}\rb.
\end{align}

For large $x$, we can replace $x=-(1+\sigma^2b^2)R$. Taking the limit $x$ large in the hypergeometric functions, $\chi_0$ simplifies to
\begin{equation}\label{chi soln x far}
\tc_0=C_1\frac{\Gamma(-\sqrt{1+4\delta})R^{-\frac{3}{2}-\frac{\sqrt{1+4\delta}}{2}}}{\Gamma\lb \frac{3}{2}-\frac{\sqrt{1+4\delta}}{2}\rb\Gamma\lb \frac{1}{2}-\frac{\sqrt{1+4\delta}}{2}\rb}+C_1\frac{\Gamma(\sqrt{1+4\delta})R^{-\frac{3}{2}+\frac{\sqrt{1+4\delta}}{2}}}{\Gamma\lb \frac{3}{2}+\frac{\sqrt{1+4\delta}}{2}\rb\Gamma\lb \frac{1}{2}+\frac{\sqrt{1+4\delta}}{2}\rb}
\end{equation}

This is the far limit of the near horizon solution which is normalizable at the horizon.
Near the singularity $r_s$ we write the equation (\ref{chi eqn near}) in terms of $y$, such that $R=m y$ and $y\sim \mathcal{O}(1)$. We denote $\chi$ as $\bc$ in this region. The singular point $r_s$ translates to $y=\frac{1}{2}$. The leading order equation is
\begin{equation}\label{chi near rs}
\frac{d^2\bc_0}{dy^2}+\lb\frac{4}{y}-\frac{4}{2y-1}\rb\frac{d\bc_0}{dy}+\frac{\lb-2\delta y-(2-\delta)\rb}{y^2(2y-1)} \bc_0=0.
\end{equation}

The general solution of this equation is
\begin{equation}\label{bc-near-rs-sol}
\bc_0=d_1y^{\frac{-3+\sqrt{1+4\delta}}{2}}((2y-1)(\sqrt{1+4\delta}-1)-2)+d_2y^{\frac{-3-\sqrt{1+4\delta}}{2}}((2y-1)(\sqrt{1+4\delta}+1)+2)
\end{equation}

Both the linearly independent solutions in (\ref{bc-near-rs-sol})  are finite at $y=\frac{1}{2}$. The solutions at $y=\frac{1}{2}$ must satisfy the condition (\ref{bdy cond at rs}). This condition rewritten in terms of $y$ simply becomes $\chi'(y)=-2\chi$ at the leading order .The condition is satisfied by any combination of $d_1$ and $d_2$. An analogous observation was made in \cite{kolstring} for $\sigma=0$. To extend this solution to the near horizon region, we rewrite $y$ in terms of $R$ and let $R\ll n$. The leading order solution then becomes,
\begin{equation}\label{chi soln y near}
\bc_0=d_1(-\sqrt{1+4\delta}-1)\lb\frac{R}{m}\rb^{\frac{-3+\sqrt{1+4\delta}}{2}}+d_2(-\sqrt{1+4\delta}+1)\lb\frac{R}{m}\rb^{\frac{-3-\sqrt{1+4\delta}}{2}}.
\end{equation}

Let us now compare the two solutions $\tc_0$ (\ref{chi soln x far}) and $\bc_0$ (\ref{chi soln y near}). The solutions coming from both sides can be matched for all values of $\lambda$ by identifying the constants $d_1$ and $d_2$ with those in (\ref{chi soln x far}). The solution at $r_s$ can then be written as
\begin{align}\label{chi soln y matched}
\bc_0&=\frac{\Gamma(\sqrt{1+4\delta})m^{\frac{-3+\sqrt{1+4\delta}}{2}}}{\Gamma\lb \frac{3}{2}+\frac{\sqrt{1+4\delta}}{2}\rb\Gamma\lb \frac{1}{2}+\frac{\sqrt{1+4\delta}}{2}\rb}\frac{y^{\frac{-3+\sqrt{1+4\delta}}{2}}((2y-1)(\sqrt{1+4\delta}-1)-2)}{(-\sqrt{1+4\delta}-1)}  \nn &+\frac{\Gamma(-\sqrt{1+4\delta})m^{\frac{-3-\sqrt{1+4\delta}}{2}}}{\Gamma\lb \frac{3}{2}-\frac{\sqrt{1+4\delta}}{2}\rb\Gamma\lb \frac{1}{2}-\frac{\sqrt{1+4\delta}}{2}\rb}\frac{y^{\frac{-3-\sqrt{1+4\delta}}{2}}((2y-1)(\sqrt{1+4\delta}+1)+2)}{(-\sqrt{1+4\delta}+1)}
\end{align}

Thus we have extended the normalizable solution near the horizon to $r_s$.
In order to match with  the solution in the far region (\ref{far soln nd}), we take  the limit $r\to \infty$ of the solution (\ref{chi soln y matched}).
\begin{align}\label{chi soln y far}
\bc_0&\approx\frac{\Gamma(\sqrt{1+4\delta})m^{\frac{-3+\sqrt{1+4\delta}}{2}}}{\Gamma\lb \frac{3}{2}+\frac{\sqrt{1+4\delta}}{2}\rb\Gamma\lb \frac{1}{2}+\frac{\sqrt{1+4\delta}}{2}\rb}\frac{R^{\frac{-1+\sqrt{1+4\delta}}{2}}}{(-\sqrt{1+4\delta}-1)}  \nn &+\frac{\Gamma(-\sqrt{1+4\delta})m^{\frac{-3-\sqrt{1+4\delta}}{2}}}{\Gamma\lb \frac{3}{2}-\frac{\sqrt{1+4\delta}}{2}\rb\Gamma\lb \frac{1}{2}-\frac{\sqrt{1+4\delta}}{2}\rb}\frac{R^{\frac{-1-\sqrt{1+4\delta}}{2}}}{(-\sqrt{1+4\delta}+1)}.
\end{align}

We can see that the solution has a growing and decaying part whereas the far region solution (\ref{far soln nd}) is only decaying. For the solutions to match, the gamma functions in the denominator of the growing piece in (\ref{chi soln y far}) must have a pole, with the numerator remaining finite. This will never be the case for coefficients of the growing part. 
We further notice that the exponents of $R$ in the two solutions take the same form only for large black holes where $1+\sigma^2b^2$ can be approximated by $\sigma^2b^2$. For the case of small black holes we cannot make a concrete statement. To do so, we would have to extend the solution in the far region to the overlap region. As in the case  $\ell >2$ (\ref{phi-far-hyp-soln}), the transformation formulae for hypergeometric functions become unwieldy. We conclude that there are no normalizable modes for $\lambda \sim \mathcal{O}(n^2)$  for the cases considered of the large black hole. This conclusion matches with the bound on the value of $\lambda$ obtained in \cite{prestidge}. Considering this result and in view of the value of $\lambda$ found in \cite{kol} for the $\sigma=0$ case i.e. the Schwarzschild-Tangherlini black holes, we shall now investigate the case $\lambda \sim \mathcal{O}(n)$.

\subsubsection{The case $\lambda \sim \mathcal{O}(n)$}

As in the previous case, we solve the equation (\ref{chi eqn near}) in the limit $R \sim \mathcal{O}(1)$ for $\tc$, and in the limit $R \sim \mathcal{O}(n)$ for $\bc$. We denote $\lambda = m L$  such that $L \sim \mathcal{O}(1)$. The expansion of $\lambda$ then becomes
\begin{equation}\label{lambdaexp}
\lambda= \sum_{i=0}^{\infty}\frac{\lambda_i}{m^i}= m L_0+ L_1 +\frac{L_2}{m}+ \dots.
\end{equation}

\noindent\textbf{(i) For $R\sim \mathcal{O}(1)$ :}
\vspace{0.1 in}

Substituting $r$ in terms of $R$ using (\ref{rR}), and expanding $\tc$ as a series in $m$, we write the equation (\ref{chi eqn near}) in orders of $m$. For $R\sim \mathcal{O}(1)$, we can approximate $\lb R-\frac{m+2}{2}\rb \approx -\frac{m}{2}$. 
 For convenience, we divide the whole equation (\ref{chi eqn near}) by $m$. Equation (\ref{chi eqn near}) then is written in orders of $m$. The leading order equation can be written as $\mathcal{L}\tc_0=0$. The  operator $\mathcal{L}$ is
\begin{equation}\label{LO operator}
\mathcal{L}=\frac{R}{2}(1-(1+\sigma^2b^2)R)\frac{d^2}{dR^2}+(1-2(1+\sigma^2b^2)R)\frac{d}{dR}-(1+\sigma^2b^2)
\end{equation}

The next order equations then are written as
\begin{equation}\label{order 1 notation}
\frac{\mathcal{L}\tc_i}{m^i}=\frac{\mathcal{S}(R)_i}{m^i}
\end{equation}
Source terms $\mathcal{S}(R)_i$ depend on the solutions of the previous order equations ($\tc_0,\dots,\tc_{i-1}$).
  The leading order equation is
\begin{equation}\label{tc LO eqn}
\frac{R}{2}(1-(1+\sigma^2b^2)R)\frac{d^2\tc_0}{dR^2}+(1-2(1+\sigma^2b^2)R)\frac{d\tc_0}{dR}-(1+\sigma^2b^2)\tc_0=0.
\end{equation}

Solutions to this equation are,
\begin{equation}
\tc_0=\frac{d_1}{R}+\frac{d_2}{(1-(1+\sigma^2b^2)R)}
\end{equation}

Normalizability at the horizon tells us $d_2=0$ so that
\begin{equation}\label{tc LO sol}
\tc_0=\frac{d_1}{R}
\end{equation}
The source term $\mathcal{S}(R)_1$ at $\mathcal{O}(1/m)$ depends on the  solution to the leading order equation $\tc_0$.
\begin{align}\label{tc NLO eqn}
\mathcal{L}\tc_1=& -\lb -3+\frac{L_0b^2}{2}-6\sigma^2b^2-2\sigma^2b^2\ln R\rb \tc_0-\nn &\lb 3-5R+2R^2-7\sigma^2b^2R+2\sigma^2b^2R^2-4\sigma^2b^2R\ln R\rb\tc_0'-\nn&
\lb R-2R^2+R^3-\sigma^2b^2R^2+\sigma^2b^2R^3-\sigma^2b^2R^2\ln R\rb \tc_0''
\end{align}

Solving for $\tc_1$, the solution is
\begin{align}\label{tc NLO sol gen}
\tc_1=&\frac{d_1^1}{R}+\frac{d_2^1}{(R(1+\sigma^2b^2)-1)}-\frac{2d_1+2d_1\ln R}{R}\nn&-\frac{(-6+b^2(L_0-4\sigma^2))d_1+(1+\sigma^2b^2)d_1^1+(2-L_0b^2)d_1\ln R}{(R(1+\sigma^2b^2)-1)}
\end{align}
Here $d_1^1, d_2^1$ are arbitrary constants corresponding to the homogeneous solution. Regularity at the horizon fixes $d_2^1$ to be
\begin{equation}
d_2^1=\lb(-6+b^2(L_0-4\sigma^2))d_1+(1+\sigma^2b^2)d_1^1-(2-L_0b^2)d_1\ln(1+\sigma^2b^2)\rb.
\end{equation}

The next to leading order solution $\tc_1$ after substituting for $d_2^1$ becomes
\begin{equation}
\tc_1=\frac{-2d_1+d_1^1-2d_1\ln R}{R}-\frac{(2-L_0 b^2)d_1(\ln (1+\sigma^2b^2)+\ln R)}{(R(1+\sigma^2b^2)-1)}
\end{equation}

We take the full solution $\tc_0+\frac{\tc_1}{m}$ to the overlap region by taking $R$ large such that $1\ll R\ll m$.
\begin{align}\label{tchi extended}
\tc&=\frac{d_1}{R}+\frac{1}{m}\left[\frac{-2d_1+d_1^1-2d_1\ln R}{R}-\right.\nn & \left.\frac{(2-L_0 b^2)d_1(\ln (1+\sigma^2b^2)+\ln R)}{R(1+\sigma^2b^2)}\lb 1+\frac{1}{R(1+\sigma^2b^2)}+\frac{1}{R^2(1+\sigma^2b^2)^2}+\dots \rb\right]
\end{align}

Here $d_1, d_1^1$ are arbitrary constants. We will match this solution to the one obtained from  the region where $R\sim\mathcal{O}(n)$. \vspace{0.1 in}

\noindent\textbf{(ii) For $R\sim \mathcal{O}(n)$ :}
\vspace{0.1 in}

We use the coordinate $y=\frac{R}{m}$  and solve for $\bc(y)$ in this region. As before, we expand $\bc$ as a series in terms of $m$. The equation (\ref{chi eqn near}) in a $1/m$ expansion in terms of $y$ can be written as
\begin{equation}
\frac{\bar{\mathcal{L}}\bc_i}{m^i}=\frac{\bar{\mathcal{S}}(y)_i}{m^i}
\end{equation}
As in equation (\ref{order 1 notation}) for $R\sim\mathcal{O}(1)$, the $\bar{\mathcal{S}}(y)_i$ are the source terms with $\bar{\mathcal{S}}(y)_0=0$.The operator $\bar{\mathcal{L}}$ is
\begin{equation}
\bar{\mathcal{L}}=(1+\sigma^2b^2)\ltb\lb y^3-\frac{y^2}{2}\rb \frac{d^2}{dy^2}+(2y^2-2y)\frac{d}{dy}-1\rtb
\end{equation}
 The leading order equation is,
\begin{equation}\label{bc LO eqn}
(1+\sigma^2b^2)\ltb\lb y^3-\frac{y^2}{2}\rb \bc_0''+(2y^2-2y)\bc_0'-\bc_0\rtb=0
\end{equation}
Solutions to this equation are
\begin{equation}\label{bc-LO-sol}
\bc_0=\frac{e_1}{y}+e_2\frac{(-1+4y^2-4y\ln y)}{y^2}
\end{equation}
where $e_1$ and $e_2$ are constants.
The equation (\ref{bdy cond at rs}) is again satisfied by both the solutions at the leading order. We can now match the solution with the behaviour of solution coming from far region (\ref{far soln dec}). This implies $e_2=0$. The leading order solution is
\begin{equation}
\bc_0=\frac{e_1}{y}
\end{equation}
At the next order,
\begin{align}
\bar{\mathcal{L}}\bc_1=&-\lb -3+\frac{b^2L_0}{2}-6\sigma^2b^2+y(-L_0b^2+2\sigma^2b^2)-2\sigma^2b^2(\ln m+\ln y)\rb \bc_0-\nn &
\lb 1+y(-5-7\sigma^2b^2-4\sigma^2b^2(\ln m+\ln y))+y^2(2+6\sigma^2b^2+4\sigma^2b^2(\ln m+\ln y))\rb \bc_0'\nn &-
\lb \frac{y}{2}+2\sigma^2b^2y^3(\ln m+\ln y)+y^2(-2-\sigma^2b^2-\sigma^2b^2(\ln m+\ln y))\rb \bc_0''
\end{align}
The solution to this equation is
\begin{align}\label{bc1 in y}
\bc_1=&\lb\frac{e_1(1-L_0 b^2-\sigma^2b^2)}{2(1+\sigma^2b^2)}-e_2^1\rb \frac{1}{y^2}+
\frac{-e_1(2+L_0 b^2+4\sigma^2b^2)+e_1^1(1+\sigma^2b^2)}{y(1+\sigma^2b^2)}\nn &+4 e_2^1+\lb-e_1(2+L_0 b^2+4\sigma^2b^2)-4e_2^1(1+\sigma^2b^2)\rb\frac{\ln y}{y(1+\sigma^2b^2)}
\end{align}
The constants $e_1^1, e_2^1$ are arbitrary constants corresponding to the solution to the homogeneous equation (added to the particular solution of the inhomogeneous equation). We want to match this solution to the near region solution. We write the solution in terms of $R$.
\begin{align}\label{bc nlo}
\bc&=\bc_0(y)+\frac{\bc_1(y)}{m}\nn
&=\frac{m e_1}{R}+\frac{m}{R^2}\lb \frac{e_1(1-L_0 b^2-\sigma^2b^2)}{2(1+\sigma^2b^2)}-e_2^1\rb  + \frac{-e_1(2+L_0 b^2+4\sigma^2b^2)+e_1^1(1+\sigma^2b^2)}{R(1+\sigma^2b^2)}\nn &+4 \frac{e_2^1}{m}+\lb-e_1(2+L_0 b^2+4\sigma^2b^2)-4e_2^1(1+\sigma^2b^2)\rb\frac{\ln R-\ln m}{R(1+\sigma^2b^2)}
\end{align}
Let us look at this solution carefully.\\
(i) We have a constant term proportional to $e_2^1$ which has no counterpart in the next to leading order solution (\ref{tchi extended}). Also, when extended to $y\approx R\to \infty$, the constant term may dominate the solution making it non-normalizable at the boundary. We hence put $e_2^1=0$.\\
(ii) Notice that in (\ref{bc nlo}), the term proportional to $(1/y^2)$ from $\bc_1$ (\ref{bc1 in y}) gets written as \begin{equation}\label{ysq term}
\frac{\bc_1}{m}\sim \frac{1}{m y^2}\sim \frac{m}{R^2}.
\end{equation}
Thus it contributes to the leading order solution after replacing $y$ as $\frac{R}{m}$ in (\ref{bc nlo}). This implies that 
 terms from the higher orders in $\bc(y)$ can contribute at the leading order in $\bc(R)$. For example upon replacing $y=\frac{R}{m}$, terms of the form $(1/m^iy^{i+1})$ at each order contribute to the leading order solution and similarly $(1/m^iy^{i})$ contribute in the next to leading order and so on.
\begin{align*}
\frac{1}{m^iy^{i+1}}\sim \frac{m}{R^{i+1}}; \quad \frac{1}{m^iy^{i}}\sim \frac{1}{R^i}
\end{align*}
To obtain the contributions from the higher order solutions to $\bc$ in (\ref{bc nlo}), we first solve for $\bc$ up to $\mathcal{O}(1/m^4)$ in terms of $y$. Then we replace $y=\frac{R}{m}$ in this solution. The full solution in $y$ is too lengthy hence we will not give it here. We will write the solution $\bc$ up to $\mathcal{O}(1/m)$ after replacing $y$ in terms of $R$. This is because we want to match the solution $\bc$ to  the solution from the near region $\tc$ obtained in (\ref{tchi extended}) which is only up to $\mathcal{O}(1/m)$.

\begin{align}\label{bc nlo full}
\bc(R)&=\frac{e_1}{R}+\frac{(-b^2 e_1 L_0-b^2 e_1 \sigma ^2+e_1)}{2 \left(b^2 \sigma ^2+1\right)}\lb \frac{1}{R^2}+\frac{1}{ R^3 \left(b^2 \sigma ^2+1\right)}+ \frac{1}{ R^4 \left(b^2 \sigma ^2+1\right)^2}\rb+ \nn
&\frac{1}{m}\left\lbrace\frac{b^2 e_1^1 \sigma ^2+b^2 e_1 L_0 \ln m-b^2 e_1 L_0+4 b^2 e_1 \sigma ^2 \ln m-4 b^2 e_1 \sigma ^2+e_1^1+2 e_1 \ln m-2 e_1}{R \left(b^2 \sigma ^2+1\right)}\right.\nn
&+\frac{\ln R \left(-b^2 e_1 L_0-4 b^2 e_1 \sigma ^2-2 e_1\right)}{R \left(b^2 \sigma ^2+1\right)}\nn
&-\frac{1}{2 R^2 \left(b^2 \sigma ^2+1\right)^2}\left[b^4 e_1^1 L_0 \sigma ^2+b^4 e_1^1 \sigma ^4-3 b^4 e_1 L_0^2-7 b^4 e_1 L_0 \sigma ^2+b^4 e_1 L_1 \sigma ^2 \right.\nn
&-3 b^4 e_1 \sigma ^4-e_1^1+3 e_1+b^2 e_1^1 L_0+b^2 e_1 L_0+b^2 e_1 L_1+4 b^2 e_1 \sigma ^2 \nn
&\left. +e_1 \ln m \left(b^4 \left(L_0^2+5 L_0 \sigma ^2+4 \sigma ^4\right)+b^2 \left(L_0-2 \sigma ^2\right)-2\right) \right] \nn
&-\frac{e_1 \ln R \left(b^4 \left(-\left(L_0^2+5 L_0 \sigma ^2+4 \sigma ^4\right)\right)+b^2 \left(L_0+6 \sigma ^2\right)+2\right)}{2 R^2 \left(b^2 \sigma ^2+1\right)^2}\nn
&-\frac{1}{4 R^3 \left(b^2 \sigma ^2+1\right)^3}\left[2 b^4 e_1^1 L_0 \sigma ^2+2 b^4 e_1^1 \sigma ^4-5 b^4 e_1 L_0^2-13 b^4 e_1 L_0 \sigma ^2+2 b^4 e_1 L_1 \sigma ^2\right.\nn
&-6 b^4 e_1 \sigma ^4+2 b^2 e_1^1 L_0+b^2 e_1 L_0+2 b^2 e_1 L_1+8 b^2 e_1 \sigma ^2 -2 e_1^1+6 e_1\nn
&\left. +2 e_1 \ln m \left(b^4 \left(L_0^2+5 L_0 \sigma ^2+4 \sigma ^4\right)+b^2 \left(L_0-2 \sigma ^2\right)-2\right)\right]\nn
&+\frac{e_1 \ln R \left(b^4 \left(L_0^2+7 L_0 \sigma ^2+6 \sigma ^4\right)-b^2 \left(L_0+8 \sigma ^2\right)-2\right)}{2 R^3 \left(b^2 \sigma ^2+1\right)^3}\nn
&-\frac{1}{6 R^4 \left(b^2 \sigma ^2+1\right)^4}\left[3 b^4 e_1^1 L_0 \sigma ^2+3 b^4 e_1^1 \sigma ^4-7 b^4 e_1 L_0^2 -19 b^4 e_1 L_0 \sigma ^2+3 b^4 e_1 L_1 \sigma ^2\right.\nn
&-9 b^4 e_1 \sigma ^4+3 b^2 e_1^1 L_0+b^2 e_1 L_0+3 b^2 e_1 L_1+12 b^2 e_1 \sigma ^2-3 e_1^1+9 e_1\nn
&\left. +3 e_1 \ln m \left(b^4 \left(L_0^2+5 L_0 \sigma ^2+4 \sigma ^4\right)+b^2 \left(L_0-2 \sigma ^2\right)-2\right)\right]\nn
&\left.+\frac{e_1 \ln R \left(b^4 \left(L_0^2 +9 L_0 \sigma ^2+8 \sigma ^4\right)-b^2 \left(L_0+10 \sigma ^2\right)-2\right)}{2 R^4 \left(b^2 \sigma ^2+1\right)^4}\right\rbrace
\end{align}

In obtaining this solution, we have used the boundary condition that $\bc_i$ is normalizable at infinity at each order.\footnote{ We have used Mathematica for detailed computations.} 
This lets us put various constants to zero similar to how we set $e_2=0$ in the leading order solution (\ref{bc-LO-sol}). 


We see that after replacing $y$ as $\frac{R}{m}$ the leading order solution acquires an $m$ factor while the leading order solution from the $R\sim \mathcal{O}(1)$ (\ref{tchi extended}) is of $\mathcal{O}(1)$. As we want to match the two solutions, we have scaled the entire solution $\bc$  by an overall factor of $m$ (recall that we had scaled the near region equations by a factor of $m$). $\bc$ solves the homogeneous equation (\ref{chi eqn near}) near $r_s$.

\noindent\textbf{(iii) Matching the solutions}

Compare the leading order terms from both $\tc$ (\ref{tchi extended}) and $\bc$ (\ref{bc nlo full}). The leading order term in (\ref{bc nlo full}) is
\begin{equation}
\bc_0(R)= \frac{e_1}{R}+\frac{(-b^2 e_1 L_0-b^2 e_1 \sigma ^2+e_1)}{2 \left(b^2 \sigma ^2+1\right)}\lb \frac{1}{R^2}+\frac{1}{ R^3 \left(b^2 \sigma ^2+1\right)}+ \frac{1}{ R^4 \left(b^2 \sigma ^2+1\right)^2}\rb
\end{equation}

The leading  order term in (\ref{tchi extended}) is $\tc_0=\frac{d_1}{R}$. The two solutions match after setting $-b^2 e_1 L_0-b^2 e_1 \sigma ^2+e_1=0$. This condition fixes $L_0$.
\begin{equation}\label{L0}
L_0b^2=1-\sigma^2b^2
\end{equation}

We are looking for the modes that have $\lambda =m L>0$. Recall (\ref{lambdaexp}). We can see that for large black holes with $\sigma b >1$, $\lambda$ becomes negative. This signifies that there are no normalizable modes for large black holes having $\lambda >0$.  For  the small black hole with $\sigma b<1$, there is a negative mode.For the Schwarzschild black hole case with $\sigma=0$, this answer matches with the leading order value of $\lambda$ obtained in \cite{kol}.

We make the coefficients of $\frac{1}{R}$ terms in both the leading order solutions equal by setting $d_1=e_1=1$.
Substituting for $L_0$ in the solutions (\ref{tchi extended}) and  (\ref{bc nlo full})  we get,
\begin{align}\label{tc final}
\tc=&\frac{1}{R}+\frac{1}{m}\ltb\frac{-2+d_1^1-\ln (1+\sigma^2b^2)}{R}-\frac{3\ln R}{R}\right. \nn
&-\ln(1+\sigma^2b^2)\lb\frac{1}{(1+\sigma^2b^2)R^2}+\frac{1}{(1+\sigma^2b^2)^2R^3}+\frac{1}{(1+\sigma^2b^2)^3R^4}\rb\nn
&\left.-\ln R\lb\frac{1}{(1+\sigma^2b^2)R^2}+\frac{1}{(1+\sigma^2b^2)^2R^3}+\frac{1}{(1+\sigma^2b^2)^3R^4}\rb \rtb.
\end{align}

We have truncated the infinite series in (\ref{tchi extended}) at $\sim R^{-4}$.  The solution $\bc$ becomes,
\begin{align}\label{bc final}
\bc=&\frac{1}{R}+\frac{1}{m}\ltb\frac{-3+e_1^1+3\ln m}{R}-\frac{3\ln R}{R}\right. \nn
&-\frac{(1+L_1b^2+\sigma^2b^2)}{2}\lb\frac{1}{(1+\sigma^2b^2)R^2}+\frac{1}{(1+\sigma^2b^2)^2R^3}+\frac{1}{(1+\sigma^2b^2)^3R^4}\rb\nn
&\left.-\ln R\lb\frac{1}{(1+\sigma^2b^2)R^2}+\frac{1}{(1+\sigma^2b^2)^2R^3}+\frac{1}{(1+\sigma^2b^2)^3R^4}\rb \rtb
\end{align}

Let us match the coefficients of the next to leading order terms. Matching $\frac{1}{R}$ terms, we get
\begin{equation}
-2+d_1^1-\ln (1+\sigma^2b^2)=-3+e_1^1+3\ln m
\end{equation}
As both $d_1^1$ and $e_1^1$ are arbitrary constants, this equation can be satisfied by choosing them to make both sides equal to zero. We see that all the terms proportional to $\ln R$ match.  The next condition comes from matching the coefficients of higher powers of $1/R$. This fixes $L_1$.
\begin{equation}\label{L1}
L_1 b^2=-1-\sigma^2b^2+2\ln (1+\sigma^2b^2)
\end{equation}

In the solution $\tc$ (\ref{tc final}), we have truncated the series in (\ref{tchi extended}) and matched the solutions. Looking at both the solutions (\ref{tc final}) and (\ref{bc final}), we expect the rest of the terms in the series in (\ref{tchi extended}) will be reproduced by calculating higher order corrections for $\bc(R)$ in (\ref{bc nlo full}). We would expect the coefficients of the terms in higher powers of $\frac{1}{(1+\sigma^2b^2)R}$ will be $-\frac{(1+L_1b^2+\sigma^2b^2)}{2}$ and that the $(\log R)$ terms will match upon substituting $L_0$ at all orders.

The value of $\lambda$ up to next to leading order is,
\begin{equation}\label{lambda final}
\lambda b^2 = m(1-\sigma^2b^2)-\ltb 1+\sigma^2b^2-2\ln (1+\sigma^2b^2)\rtb.
\end{equation}

For $\sigma=0$, i.e. the Schwarzschild black hole case, this value of $\lambda$ matches that obtained in \cite{kol} till next to leading order. In fact for small black holes taking $\sigma b \ll 1$, we can write our eigenvalue
\begin{equation}
\lambda b^2=(m-1)(1-\sigma^2b^2).
\end{equation}

We have calculated $\lambda$ only up to two orders here. We would need to calculate higher order solutions of both $\tc$ and $\bc$ in order to calculate higher order corrections to $\lambda$. The solution for $\tc$ at the next to leading order contains polylogarithmic functions in $R$. The solutions then become very hard to analyse. Hence we have not proceeded further.

\section{Results and Discussion}\label{results}

In this paper we have studied the semiclassical stability problem for the SAdS black hole under scalar perturbations. We obtain the equations for non-spherically symmetric perturbations using the gauge invariant formalism of IK \cite{iks},\cite{vac}. These equations are a generalization to the case of nonzero cosmological constant from the equations obtained in \cite{noneg} for the Schwarzschild-Tangherlini black hole. One of the three resulting equations decouples from the others. We analyze the decoupled equation using the S-deformation approach \cite{vac}, \cite{wald}. We demonstrate semiclassical stability for this class of perturbations. We note that while in the rest of the paper we have used the large $D$ limit as an analytical approximation tool, for this  class of perturbations, we did not use this approximation. Further, this result is valid for both large and small black holes.

We next analyze the two coupled equations in the non-spherically symmetric perturbation sector. This is done using the large $D$ limit.
In the large $D$ limit we have two naturally overlapping regions, a near and a far region.  We find the appropriate solution in each region and extend each of them to the overlap region to attempt matching of solutions. For a class of perturbations we prove the absence of unstable modes with eigenvalue of $\mathcal{O}(D^2)$ in the case of the large black hole. While the near region analysis is possible for the small black hole, the far region solution cannot be continued to a simple expression in the overlap region.

We finally analyze the spherically symmetric perturbation using the large $D$ approach. We analyze the equation in the near and far regions as before. The equation in the near region is a Heun equation making it difficult to find solutions except locally around a singular point. There are two singular points in the near region, one being the horizon where $R\sim \mathcal{O}(1)$ and the other at $R\sim \mathcal{O}(D)$. We employ a procedure first used in \cite{est} to find decoupled quasinormal modes of Schwarzschild-Tangherlini black holes in the large $D$ limit. We find the normalizable solution  for the subregion with $R\sim \mathcal{O}(1)$ and the  general solution in the subregion with $R\sim \mathcal{O}(D)$, then extend them to an overlap region $1\ll R\ll D$ and match the two solutions. After this matching procedure, the solution around the second singular point is compared with the normalizable solution in the far region. We find that there is no unstable mode with eigenvalue of $\mathcal{O}(D^2)$. This is consistent with the bound on this eigenvalue obtained by Prestidge \cite{prestidge}. On the other hand, there is indeed an unstable mode with eigenvalue of $\mathcal{O}(D)$ for the small black hole which disappears on increasing the mass. Thus the large black hole is stable. This is the analogue of the Hawking-Page phase transition in semiclassical stability analysis. A delicate calculation in a $1/D$ expansion is needed to obtain the eigenvalue corresponding to this unstable mode. We obtain this eigenvalue for the first time in a large $D$ limit. We compute the eigenvalue to next to leading order and for $\Lambda =0$, this agrees with the value of the unstable mode in \cite{kol} for the Schwarzschild black hole in the large $D$ limit.  The agreement is to next to leading order. As shown in \cite{kol}, the $\Lambda=0$ value corresponds to the Gross-Perry-Yaffe unstable mode \cite{gpy} for the $D$-dimensional Schwarzschild instanton which is also mathematically related to the Gregory-Laflamme unstable mode \cite{GL} for the $(D+1)$-dimensional flat black string. The significance of the unstable mode is that there exists a metric perturbation of the small black hole that decreases the Euclidean action. Thus, in the Euclidean path integral approach to quantum gravity, the small black hole is a saddle point of the action, not a true minimum.

 \appendix
 \setcounter{section}{0}
 \setcounter{subsection}{0}
 \setcounter{equation}{0}
 \renewcommand\theequation{\Alph{section}\raise.5ex\hbox{.}\arabic{equation}}

\section{Appendix A: Equations for Scalar Perturbations}\label{AppA}

\renewcommand\theequation{\Roman{equation}}
In this section we give an outline of the procedure to get equations in terms of the $W,Y,Z$ variables. We first write equations $2\delta G_{\mu\nu}+2\Lambda h_{\mu\nu}=-\lambda  h_{\mu\nu}$ in terms of the gauge invariant variables $(F_{ab},F)$ (expressions for $2\delta G_{\mu\nu}+2\Lambda h_{\mu\nu}$ in terms of these variables are given in appendix A in \cite{vac}) . We then change the variables to $W,Y$ and $Z$ using following expressions:
\begin{align*}
&F=-\frac{W+Y}{2n r^{n-2}}+\frac{\lambda}{nk^2}(r^2H_T) \qquad &&F^r_t=\frac{Z}{r^{n-2}} \\
&F^t_t=\frac{W(n-1)-Y}{2n r^{n-2}}+\frac{2\lambda}{nk^2}(r^2H_T) \qquad &&F^r_r=\frac{Y(n-1)-W}{2n r^{n-2}}+\frac{2\lambda}{nk^2}(r^2H_T).
\end{align*}

Equation for $\delta G_{ti}$ :
\begin{equation}\label{f_t}
\partial_t W + \partial_r Z = \lambda  r^{n-2} \left[X_t +\frac{1}{k^2}\partial_t(r^2H_T)\right]
\end{equation}

Equation for $\delta G_{ri}$ :
\begin{equation}\label{f_r}
\partial_r Y+\frac{f'}{2f}Y-\frac{f'}{2f}W -\frac{1}{f^2}\partial^2_t Z = \lambda r^{n-2}\left[X_r+\frac{1}{k^2}\p_r(r^2H_T)\right]
\end{equation}

Equation for $\delta G_{t}^r$ :
\begin{align}\label{f_rt}
&\left[\frac{k^2}{r^2}-f'' -\frac{nf'}{r}-\frac{4\Lambda}{n} \right]Z
    + f\partial_t\partial_r Y
   + \left(\frac{2f}{r}-\frac{f'}{2}\right) \partial_t Y  \nonumber \\
 &+ f\partial_t\partial_r W
  - \left(\frac{(n-2)f}{r}+\frac{f'}{2}\right) \partial_t W =-\lambda  r^{n-2}f^r_t \nonumber \\& - \frac{2\lambda }{nk^2}\left[n\p_t\p_r(r^2H_T)-\frac{nf'}{r}\p_t(r^2H_T)\right]
\end{align}

Equation for $\delta G_{r}^r$ :
\begin{align}\label{f_rr}
& \frac{1}{f}\partial_t^2 W -\frac{f'}{2}\partial_r W +\frac{1}{f}\partial_t^2 Y -\left(\frac{f'}{2}+\frac{nf}{r}\right)
   \partial_r Y \notag\\
& +\left[\frac{n-1}{r^2}(f-1)+\frac{4\Lambda}{n^2}+\frac{(n+2)f'}{2r}+\frac{f''}{n}\right]W \notag\\
& +\left[\frac{1-f}{r^2}-\frac{4(n-1)\Lambda}{n^2}-\frac{3n-2}{2r}f'-\frac{n-1}{n}f''+\frac{k^2-nK}{r^2}\right]Y
    \notag\\
& + \frac{2n}{rf}\partial_t Z = -\lambda  r^{n-2}f^r_r \notag \\
& -\frac{2\lambda }{nk^2}\left[-\frac{nk^2}{r^2}(H_Tr^2)-\frac{n}{f}\p^2_t(r^2H_T)+\frac{nf'}{2}\p_r(r^2H_T)+\frac{n^2f}{r}\p_r(r^2H_T)+2\Lambda(r^2H_T)\right]
\end{align}

Equation for $\delta G_i^i$ :
\begin{align}\label{f_L}
& \frac{1}{2f} \partial_t^2 W +\frac{f'}{4} \partial_r W -\frac{f}{2} \partial_r^2 Y
   -\left(\frac{3f'}{4}+\frac{f}{r} \right)\partial_r Y \notag\\
& +\left[ \frac{(n-1)(n-2)(f-1)}{2nr^2}
   +\frac{(6n-4-n^2)f'}{4nr} +\frac{f''}{2n} \right] W \notag\\
& +\left[ \frac{(n-1)(n-2)(f-1)}{2nr^2}
    +\frac{(-n^2+2n-4)f'}{4nr}-\frac{(n-1)f''}{2n}\right] Y \notag\\
& +\left(\frac{1}{rf}-\frac{f'}{2f^2}\right) \partial_t Z
   +\frac{1}{f}\partial_t\partial_r Z=-\lambda  r^{n-2}H_L +\frac{\lambda }{nk^2}\bigg[\frac{(n-1)k^2}{r^2}(H_Tr^2)\nn &-nf\p^2_r(r^2H_T)+\frac{n}{f}\p^2_t(r^2H_T)-nf'\p_r(r^2H_T)-\frac{n(n-1)f}{r}\p_r(r^2H_T)-2\Lambda(r^2H_T) \bigg]
\end{align}

Equation for $\delta G_{t}^t$ :
\begin{align}\label{f_tt}
 & -f\partial_r^2 W+ \left(\frac{n-4}{r}f-\frac{f'}{2}\right)
    \partial_r W -f\partial_r^2 Y - \left(\frac{f'}{2}+\frac{4f}{r}\right)
    \partial_r Y  \notag\\
& -\left[\frac{n-1}{r^2}-\frac{(2n-3)f}{r^2}+\frac{4(n-1)\Lambda}{n^2}
    +\frac{n-2}{2r}f'+\frac{n-1}{n}f''
    -\frac{k^2}{r^2}\right]W \notag\\
& -\left[\frac{n-1}{r^2}-\frac{4\Lambda}{n^2}-\frac{n-3}{r^2}f
   + \frac{(n-2)f'}{2r}
   -\frac{f''}{n}\right] Y =-\lambda  r^{n-2}f^t_t \notag \\
&  -\frac{2\lambda }{nk^2}\left[-\frac{nk^2}{r^2}(H_Tr^2)+nf\p^2_r(r^2H_T)+\frac{nf'}{2}\p_r(r^2H_T)+\frac{n^2f}{r}\p_r(r^2H_T)+2\Lambda(r^2H_T)\right]
\end{align}

As discussed before, the right-hand side of these equations have components of $(f_{ab},X_a,H_L,H_T)$. To get them in terms of $W,Y,Z$, we have to combine these equations. Let us expand our variables in terms of these components.

\begin{align*}
& \frac{W}{r^{n-2}} = f^t_t-\frac{2}{f}\p_tX_t+\left(f'-\frac{2f}{r}\right)X_r-2H_L-\frac{2H_T}{n}\\
& \frac{Y}{r^{n-2}} = f^r_r+2f\p_rX_r+\left(f'-\frac{2f}{r}\right)X_r-2H_L-\frac{2H_T}{n}	\\
& \frac{Z}{r^{n-2}} = f^r_t+f\p_rX_t+f\p_tX_r-f'X_t
\end{align*}

\setcounter{equation}{0}
\renewcommand\theequation{\alph{equation}}
The final $W,Y$ and $Z$ equations are obtained as follows:
Looking at the expression of $Z$, we see that adding equation (\ref{f_rt}) and derivatives of equations (\ref{f_t}) and (\ref{f_r}) with appropriate coefficients will give the right-hand side of the resulting equation in terms of the $Z$ variable.
Similarly adding equation (\ref{f_rr}) to the derivatives of equations (\ref{f_r}) and (\ref{f_L}) will give us the equation for $Y$.
To obtain the equation for $W$, we add equation (\ref{f_tt}), the derivative of equation (\ref{f_t}), equations (\ref{f_L}) and (\ref{f_r}).
We see that in this process, the extra $H_T$ terms get cancelled in the resulting equations.  In the static limit, the $Z$ equation decouples and we get coupled equations for $(W,Y)$. We then rewrite these equations  $\phi,\psi$ and $\eta$ variables defined in (\ref{Sca-def-sifi})

\setcounter{equation}{0}
\section{Appendix B: Solutions to the $\eta$ equation}\label{AppB}

In this section we shall obtain solutions to the $\eta$ equation
that are normalizable at the boundaries. We have defined $\xi=f^{-\frac{1}{2}}\eta$. The equation for $\xi$ is written in Schrodinger form in coordinate $r_*$ with potential
\begin{equation}\label{xisol1}
V=\frac{(n^2-2n)}{4r^2}f^2-\frac{(n+2)}{2}\frac{f'f}{r}+f'^2-2f''f+\frac{k^2}{r^2}f+\lambda f+2(n+1)\sigma^2f
\end{equation}

Recall $dr_*=\frac{dr}{f(r)}$ and $f(r)=1+\sigma^2r^2-\frac{b^{n-1}}{r^{n-1}}$. We cannot fully integrate $dr_*$ to find the tortoise coordinate $r_*$. Instead we shall find $r_*$ in terms of $r$ near the horizon $(r=r_+)$ and infinity.

Very near the horizon, we can approximate
\begin{equation}\label{xisol2}
f(r)\approx f'(r_+)(r-r_+)=\frac{(n+1)\sigma^2r_+^2-2}{r_+}(r-r_+)=\alpha (r-r_+)
\end{equation}
Consequently, very near the horizon $r_*$ becomes
\begin{equation}\label{xisol3}
r_*=\int\frac{dr}{f(r)}\approx \frac{\ln (r-r_+)}{\alpha}
\end{equation}

In this limit as $r\to r_+$, the coordinate $r_*\to -\infty$.
Therefore, very near the horizon the potential (\ref{xisol1}) can be approximated by constant $v(r)\approx f'(r_+)^2=\alpha^2$. The Schrodinger equation for $\xi$ in this limit is,
\begin{equation}\label{xisol4}
-\frac{d^2\xi}{d r_*^2}+\alpha^2\xi=0
\end{equation}
Solutions to this equation are $\xi = c_1 e^{\alpha r_*}+c_2 e^{-\alpha r_*}$. As $r_*\to -\infty$, the second solution $e^{-\alpha r_*}\to \infty$ whereas $e^{\alpha r_*} \to 0$. From  the normalizable solution $\xi$ at the horizon is
\begin{equation}\label{xisol5}
\xi=c_1 e^{\alpha r_*}=c_1(r-r_+)
\end{equation}

Similarly we can find $r_*$ near the infinity. In this limit $r_*=\frac{1}{\sigma r_*}$. As $r\to \infty$, $r_*\to 0$. Substituting $f(r)$ in the potential (\ref{xisol1}) equation for $\xi$ near $r\to \infty$ becomes
\begin{equation}\label{xisol6}
-\frac{d^2\xi}{d r_*^2}+\lb \frac{n^2}{4}+\frac{n}{2}+\frac{\lambda}{\sigma^2}\rb \frac{1}{r_*^2}\xi=0
\end{equation}
Solutions to this equation are
\begin{equation}\label{xisol7}
\xi= d_1 r_*^{\frac{1}{2}+\frac{1}{2}\sqrt{(n+1)^2+\frac{4\lambda}{\sigma^2}}}+d_2 r_*^{\frac{1}{2}-\frac{1}{2}\sqrt{(n+1)^2+\frac{4\lambda}{\sigma^2}}}
\end{equation}
For $\lambda >0$ and $n\geq 2$ the quantity $(n+1)^2+\frac{4\lambda}{\sigma^2} >1$. The normalizable solution at $r_*=0$ is thus obtained by $d_2=0$. The normalizable solution $\xi \to 0$ as $r\to \infty$.
\begin{equation}\label{xisol8}
\xi=d_1 r_*^{\frac{1}{2}+\frac{1}{2}\sqrt{(n+1)^2+\frac{4\lambda}{\sigma^2}}}
\end{equation}

To show stability of spacetime under $\eta$ perturbations, we have used $S$-deformation method and showed that the deformed potential (\ref{xi-def-pot}) is positive for $S$ given by (\ref{S-deformation}).
To arrive at this result we had assumed that the boundary terms in (\ref{xi-Sintegrated}) vanish for our chosen $S$. From the solutions obtained near the horizon (\ref{xisol5}) and  near the infinity (\ref{xisol8}), we see that for $S=f'(r)$, the term $S|\xi|^2$ in (\ref{xi-Sintegrated}) vanishes at both the boundaries. The other boundary term in (\ref{xi-Sintegrated}) also vanishes as $\xi \to 0$ at both the boundaries.

\setcounter{equation}{0}
\section{Appendix C: Expansion of Modified Bessel Functions}\label{AppC}

In this section, we write the expansion of $I_\nu\lb\frac{k}{\sigma r} \rb$ in terms of $R$. Recall, $\nu=\sqrt{\frac{n^2+1}{4}+\frac{\lambda}{\sigma^2}}$, $k^2=\ell(\ell+n-1)$.

In the modified Bessel function $I_\nu\lb\frac{k}{\sigma r} \rb$, as we are working with $k^2,\lambda \sim \mathcal{O}(n^2)$ , both the Bessel function order $\nu$ and argument $\lb\frac{k}{\sigma r} \rb$ are $\mathcal{O}(n)$. As we are working in the large $n$ approximation, we have to use
expansions for modified Bessel functions of large order and large argument.

Let us denote $\kappa=\frac{k}{\sigma}$. For simplicity of calculation, we define a new coordinate $z=\frac{\kappa}{\nu r} $.  Here
\begin{equation}
\nu=\sqrt{\frac{n^2+1}{4}+\frac{\lambda}{\sigma^2}}\approx\frac{n}{2}\sqrt{1+4\frac{\hat{\lambda}}{\sigma^2}}
\end{equation}

The far region solution (\ref{far-phi-sol}) can be now written as
\begin{equation*}
\phi=D_2 \frac{I_\nu(\nu z)}{\sqrt{z}}
\end{equation*}

The large order and large argument expansion of this expression is,
\begin{equation}
\frac{c_1I_\nu(\nu z)}{\sqrt{z}}=\frac{1}{\sqrt{z}}\frac{e^{\nu\eta}}{(1+z^2)^{1/4}\sqrt{2\pi\nu}}\left[1+\sum_{m=1}^{\infty}\frac{U_m(\tilde{t})}{\nu^m}\right]
\end{equation}

where
\begin{eqnarray}
\eta &=& \sqrt{1+z^2}+\ln\left[\frac{z}{1+\sqrt{1+z^2}}\right]; \nonumber \\
\tilde t &=& \frac{1}{\sqrt{1+z^2}}.
\end{eqnarray}

and $U_m(\tilde t)$ are polynomials in $\tilde t$. We are only considering terms to highest order in $n$. We ignore the polynomial terms as they are divided by $\nu$. Substituting for $\eta$ we get, up to a constant,

\begin{equation}\label{z_exp}
\frac{I_\nu(\nu z)}{\sqrt{z}}\approx\frac{z^\nu}{\left(1+\sqrt{1+z^2}\right)^\nu}\frac{1}{\sqrt{z}(1+z^2)^{1/4}}\exp\left[\nu\sqrt{1+z^2}\right].
\end{equation}

In order to express $I_\nu(\nu z)$ in the overlap region, we write this expression in terms of $R$. Here, $R=\frac{r^{n-1}}{b^{n-1}}$. To expand the expression in orders of $n$, we use the following definition of $r$ in terms of $R$, valid in the overlap region,
\begin{equation*}
r=b \left[1+\frac{\ln R}{n-1}\right].
\end{equation*}

We will now look at each term in (\ref{z_exp}) individually. For $z^\nu$, the term is directly proportional to $r^{-n}$. Hence we use the definition $r^n=b^n R$ for large $n$.
\begin{equation}
z^\nu=\left(\frac{\kappa}{\nu}\right)^\nu \frac{1}{r^\nu}=\left(\frac{\kappa}{\nu b}\right)^\nu R^{-\frac{\nu}{n}}=\left(\frac{\kappa}{\nu b}\right)^\nu\exp\ltb -\frac{\nu \ln R}{n}\rtb
\end{equation}

The next term becomes

\begin{align}
&\left[1+\sqrt{1+z^2}\right]^{-\nu}=\exp\left[-\nu\ln\left\lbrace 1+ \lb 1 + \frac{\kappa^2}{\nu^2b^2}\lb 1-2\frac{\ln R}{n}\rb\rb^{1/2}\right\rbrace\right]\nn
&= \lb1+\sqrt{1+\frac{\kappa^2}{\nu^2b^2}}\rb^{-\nu} \exp\ltb \frac{\kappa^2}{\nu b^2 n\sqrt{1+\frac{\kappa^2b^2}{\nu^2}} \lb 1+\sqrt{1+\frac{\kappa^2b^2}{\nu^2}}\rb} \ln R\rtb .
\end{align}

$\kappa$ and $\nu$ are of order $n$. Hence the constant multiplying $\ln R$ is of  $\mathcal{O}(1)$.
Similarly substituting for $z$ we get,
\begin{equation}
\exp\ltb \nu\sqrt{1+z^2}\rtb = \lb 1+\frac{\kappa^2}{\nu^2b^2}\rb^{\nu/2}\exp\ltb \frac{-\kappa^2}{n\nu b^2\sqrt{1+\frac{\kappa^2}{\nu^2b^2}}}\ln R\rtb
\end{equation}

The remaining term in (\ref{z_exp}) becomes,
\begin{align}
\frac{1}{\sqrt{z}(1+z^2)^{1/4}}= \lb\frac{\nu b}{\kappa}\rb^{1/2}\lb 1+ \frac{\kappa^2}{\nu^2b^2}\rb^{-1/4}\ltb 1+\frac{1+2\frac{\kappa^2}{\nu^2b^2}}{1+\frac{\kappa^2}{\nu^2b^2}}\frac{\ln R}{n}\rtb
\end{align}

Here, the constant multiplying $\ln R$ is of order $1/n$. Therefore, this term is sub-leading in comparison with the other terms in expansion. We are interested in terms that are leading order in $n$. In the final expression, we neglect this term. Substituting all the expressions in $R$ back in (\ref{z_exp}), we get the following expression for $\phi$ (we have absorbed all the constants in $D_0$) :

\begin{equation}
\phi=D_0 R^{-\frac{1}{2}\sqrt{1+\frac{4(\hat{\lambda}b^2+\hat{k}^2)}{\sigma^2b^2}}}
\end{equation}

Expansion for the modified Bessel function of the second kind $K_\nu(\nu z)$ for large order and large argument is
\begin{equation}
K_{\nu}(\nu z)=\frac{\sqrt{\pi}}{\sqrt{2\nu}}\frac{1}{(1+z^2)^{1/4}}e^{-\nu\eta}\left[1+\sum_{m=1}^{\infty}(-1)^m\frac{U_m(\tilde{t})}{\nu^m}\right]
\end{equation}

We can obtain the expansion of $K_\nu(\nu z)$ in the overlap region in the large $n$ approximation by replacing $\nu$ by $(-\nu)$ in the expansion of $I_\nu(\nu z)$. The final expression for $I_\nu$ is
\begin{equation}\label{I_large_n}
K_\nu(\nu z)= (const.)R^{\frac{1}{2}\sqrt{1+\frac{4(\hat{\lambda}b^2+\hat{k}^2)}{\sigma^2b^2}}}
\end{equation}

\section{Appendix D: Handling source terms}\label{AppD}

 In the far region calculation, we have considered the case $\psi\sim r^\delta \psi$ with $\delta = 2$ or $\delta<1$, where the equation for $\phi$ decouples.
To get the solution for $\psi$ in the far region , we need to solve the inhomogeneous equation (\ref{psi-far-bes-eqn}) with $\phi$ as a source term. Considering the coupling terms to leading order, the $\psi$ equation becomes
\begin{equation}\label{sourcea1}
-\frac{d^2\psi}{dr^2}+\ltb \lb \frac{\lambda}{\sigma^2}+\frac{n^2}{4}\rb\frac{1}{r^2}+\frac{k^2}{\sigma^2r^4}\rtb\psi=\frac{2}{\sigma^2r^2}\phi.
\end{equation}

To find solutions of this equation, we first change the coordinates to $x=\frac{k}{\sigma r}$. From (\ref{far-phi-sol}), we know $\phi=D\frac{I_\nu(x)}{\sqrt{x}}$, where $\nu=\sqrt{\frac{n^2+1}{4}+\frac{\lambda}{\sigma^2}}$. We also write $\psi=x^{-1/2}M$. The equation (\ref{sourcea1}) then becomes
\begin{equation}\label{sourcea2}
x^2\frac{d^2M}{dx^2}+x\frac{dM}{dx}-(\nu^2+x^2)M=\frac{2}{\sigma^2}D_0I_\nu(x)
\end{equation}

The Wronskian of the two linearly independent solutions to the homogeneous equation
$W[I_\nu (x), K_\nu (x)] = \frac{1}{x}$. We use the method of variation of parameters to write the solution to
(\ref{sourcea2}). This takes the form
\begin{equation}
S = - \frac{2}{\sigma^2} I_\nu (x) \int K_\nu (x)I_\nu(x) x ~dx  ~+~
\frac{2}{\sigma^2} K_\nu (x) \int (I_\nu (x))^2 x ~dx.
\label{sourcea3}
\end{equation}

To evaluate this expression in the overlap region, we replace $x=\frac{k}{\sigma b}R^{-1/n}$. From appendix B, we know the expressions for $I_\nu(x)=I_\nu\lb \frac{k}{\sigma r}\rb=c R^{-\Delta}$ and $K_\nu(x)=\tilde{c} R^\Delta$ where $\Delta={\frac{1}{2}\sqrt{1+\frac{4(\hat{\lambda}b^2+\hat{k}^2)}{\sigma^2b^2}}}$. We have used the uniform asymptotic expansions of the modified Bessel functions to get these expressions. From these expansions, note that $I_\nu(x)K_\nu(x) \sim \frac{1}{\nu}$.  Evaluating the integrals we see that $S \approx (const.)R^{-\Delta}$. Consequently we write
\begin{equation}
\psi=(const.)R^{{-\frac{1}{2}\sqrt{1+\frac{4(\hat{\lambda}b^2+\hat{k}^2)}{\sigma^2b^2}}}}.
\end{equation}

\end{document}